\documentclass[journal,10pt]{IEEEtran}
\usepackage{algpseudocode}                                 
\usepackage{algorithm}
\usepackage{graphicx}                                      
\usepackage{amsmath}
\usepackage{amssymb}
\usepackage{amsfonts}
\usepackage{booktabs}
\usepackage{multirow}
\usepackage[subrefformat=parens,farskip=0pt,justification=centering]{subfig}
\usepackage{color}
\usepackage{cite}                                          
\usepackage{comment}                                       
\usepackage{soul}                                          
\soulregister\cite7
\soulregister\ref7
\soulregister\pageref7
\usepackage{amsthm}
\usepackage{etoolbox}                                      
\usepackage{url}
\usepackage{nth}                                           
\usepackage{bm}                                            
\usepackage{balance}
\usepackage{threeparttable}
\usepackage[bookmarks=false]{hyperref}                     %
\hypersetup{
    colorlinks = true,
    citecolor  = blue,
    linkcolor  = blue,
    urlcolor   = blue,
}
\usepackage{filecontents}                                  
\usepackage{pgfplots}
\usepackage{pgfplotstable}
\pgfplotsset{compat=newest}
\algrenewcommand\textproc{\texttt}

\makeatletter
\let\OldStatex\Statex
\renewcommand{\Statex}[1][3]{%
  \setlength\@tempdima{\algorithmicindent}%
  \OldStatex\hskip\dimexpr#1\@tempdima\relax
}
\makeatother

\RequirePackage[normalem]{ulem} 
\RequirePackage{color}\definecolor{RED}{rgb}{1,0,0}\definecolor{BLUE}{rgb}{0,0,1} 

\renewcommand{\vec}[1]{\boldsymbol{#1}}

\newcommand{\eqRef}[1]{Eq.~\eqref{#1}}
\newcommand{\tabRef}[1]{TABLE~\ref{#1}}
\newcommand{\figRef}[1]{Fig.~\ref{#1}}
\newcommand{\secRef}[1]{Section~\ref{#1}}
\newcommand{\algRef}[1]{Algorithm~\ref{#1}}

\algrenewcommand\algorithmicrequire{\textbf{Input:}}
\algrenewcommand\algorithmicensure{\textbf{Output:}}

\begin{document}
	
\title{LEAPS: Topological-\underline{L}ayout-Adaptable Multi-Di\underline{e} FPG\underline{A} \underline{P}lacement for \underline{S}uper Long Line Minimization}

\author{Zhixiong Di,~\IEEEmembership{Member,~IEEE}, 
		Runzhe Tao,
		Jing Mai,
		Lin Chen,
		Yibo Lin,~\IEEEmembership{Member, ~IEEE} 
		
	\thanks{This work was supported by National Natural Science Foundation of China (62374138, 62034007). Corresponding author: Zhixiong Di (dizhixiong2@126.com).
		
		Zhixiong Di, Runzhe Tao, and Lin Chen are with the School of Information Science and Technology, Southwest Jiaotong University, Chengdu, China. (e-mail:dizhixiong2@126.com, 825140517@qq.com, mix\_lc@qq.com). 
		
		Jing Mai is with the School of Computer Science and the School of Integrated Circuits, Peking University, Beijing, China. (email: jingmai@pku.edu.cn)
		
		Yibo Lin is with the School of Integrated Circuits, Peking University, Beijing, China, Institute of Electronic Design Automation, Peking University, Wuxi, China, and Beijing Advanced Innovation Center for Integrated Circuits, Beijing, China. (email: yibolin@pku.edu.cn)
	}
}

\maketitle

\begin{abstract}
Multi-die FPGAs are crucial components in modern computing systems, particularly for high-performance applications such as artificial intelligence and data centers. 
Super long lines (SLLs) provide interconnections between super logic regions (SLRs) for a multi-die FPGA on a silicon interposer. 
They have significantly higher delay compared to regular interconnects, which need to be minimized. 
With the increase in design complexity, the growth of SLLs gives rise to challenges in timing and power closure. 
Existing placement algorithms focus on optimizing the number of SLLs but often face limitations due to specific topologies of SLRs. 
Furthermore, they fall short of achieving continuous optimization of SLLs throughout the entire placement process. 
This highlights the necessity for more advanced and adaptable solutions.

In this paper, we propose \texttt{LEAPS}, a comprehensive, systematic, and adaptable multi-die FPGA placement algorithm for SLL minimization. 
Our contributions are threefold: 
1) proposing a high-performance global placement algorithm for multi-die FPGAs that optimizes the number of SLLs while addressing other essential design constraints such as wirelength, routability, and clock routing; 
2) introducing a versatile method for more complex SLR topologies of multi-die FPGAs, surpassing the limitations of existing approaches; and 
3) executing continuous optimization of SLL counts across the whole placement stages, including global placement (GP), legalization (LG), and detailed placement (DP). 
Experimental results demonstrate the effectiveness of \texttt{LEAPS} in reducing SLLs and enhancing circuit performance. 
Compared with the most recent state-of-the-art (SOTA) method, \texttt{LEAPS} achieves an average reduction of 43.08\% in SLL counts and 9.99\% in HPWL while exhibiting a notable 34.34$\times$ improvement in runtime.
\end{abstract}


\section{Introduction}
\begin{figure}
	\centering
	\includegraphics[width=\columnwidth]{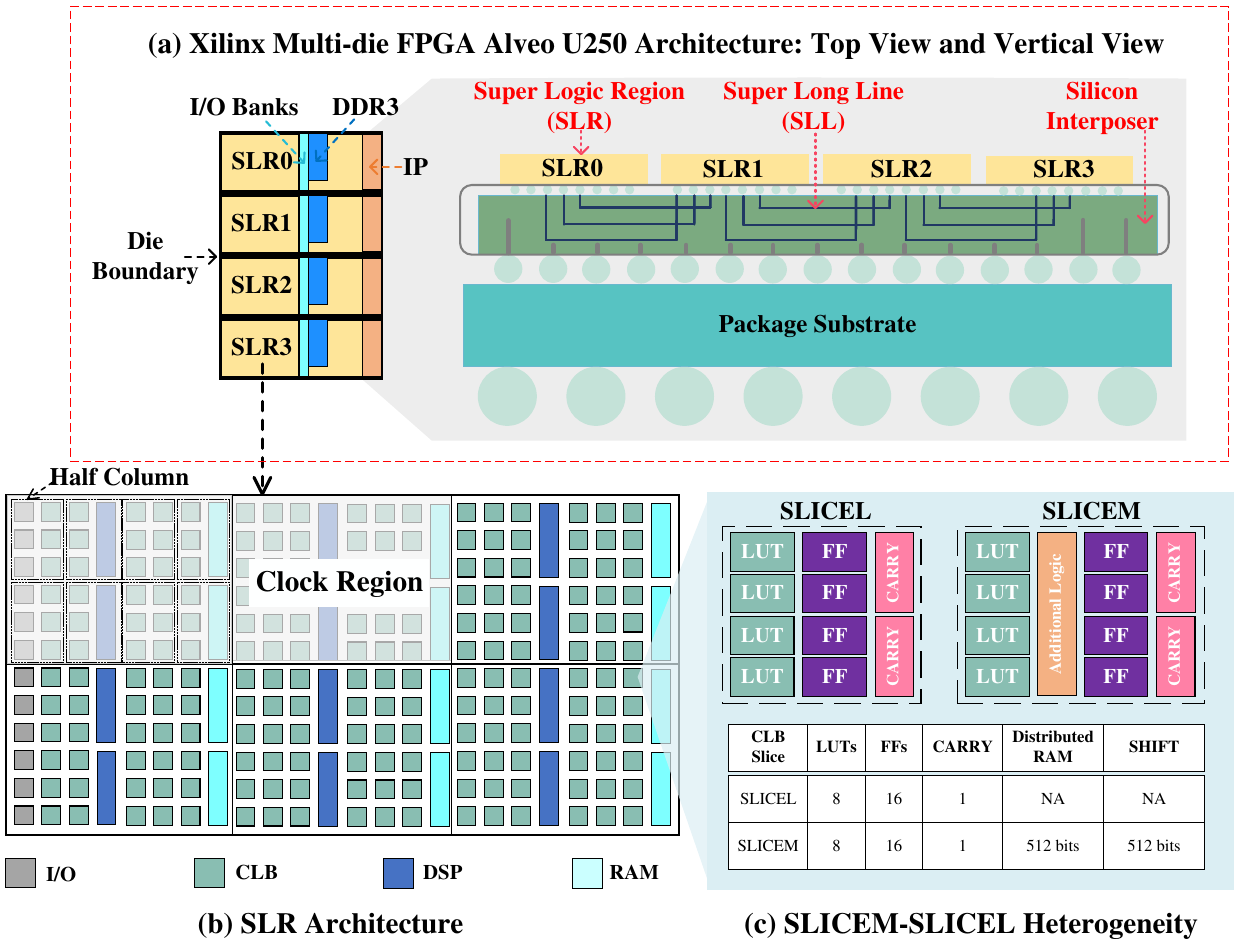}
	\caption{
		(a) Architectural illustration of Xilinx multi-die FPGA Alveo U250: Demonstrating a $1 \times 4$ SLR topology with central I/O banks and DDR controller IPs, and a right-side Vitis platform for CPU communication. 
		(b) Detailed view of SLR architecture: Partitioned into $2 \times 3$ clock regions and further segmented into multiple half columns.  
		(c) Schematic of a CLB slice: Distinguishing between SLICEL and SLICEM types to highlight asymmetric compatibility. 
	}
	\label{fig:Multi-die FPGA Arch.}
\end{figure}

\IEEEPARstart{M}{ulti-die} FPGAs are essential for modern computing systems, especially for high-performance applications such as artificial intelligence and data centers. 
A multi-die FPGA comprises several SLRs on a silicon interposer, interconnected by SLLs that facilitate communication between these regions, as depicted in \figRef{fig:Multi-die FPGA Arch.}\textcolor{blue}{(a)}. 
In the multi-die FPGA design flow, cells within each SLR are interconnected by routing resources (i.e. regular interconnects), while SLLs enable interconnections between SLRs.
Nevertheless, it is essential to emphasize that SLLs exhibit substantially greater latency in comparison to regular interconnects, consequently severely impacting timing performance.
As design complexity grows, the number of SLLs multiplies, which leads to performance degradation and power increase.
Therefore, minimizing the SLL counts is a crucial and challenging task in multi-die FPGA placement.

Existing works~\cite{pin_assign_1, pin_assign_2, modular_place_1, modular_place_2, analytical_place_3d_poisson, place_survey, vlsi_placement_research} have endeavored to optimize the number of SLLs during partitioning before placement.
In~\cite{pin_assign_1} and~\cite{pin_assign_2}, they solve the SLL issue by employing distinct optimization techniques for solving the pin assignment problem. 
Specifically, the former utilizes integer linear programming (ILP), while the latter combines a cluster approach with minimum
cost flow (MCF) optimization.
Moreover, modular placement approaches have been investigated in~\cite{modular_place_1} and~\cite{modular_place_2}, where optimal SLL resource utilization is achieved by mapping partitioned modules to appropriate dies. 
However, these approaches cannot simultaneously consider various physical constraints like clock routing, as they are applied at a separate partitioning stage before placement.
A recent state-of-the-art (SOTA) approach~\cite{analytical_place_3d_poisson} proposes an analytical placement method for multi-die FPGAs, 
which optimizes both the number of SLLs and critical clock routing constraints based on a 3D Poisson density formulation with proximal alternating direction method of multipliers (ADMM) as the solver. However, this method is
only applicable to a specific multi-die FPGA architecture (e.g., four dies arranged vertically on an interposer) and cannot accommodate more complex topologies.

Additionally, these established placement methods typically concentrate on SLL optimization during the global placement (GP) stage, often neglecting the necessity to tackle the SLL issue in the subsequent legalization (LG) and detailed placement (DP) stages.
This oversight may inadvertently degrade circuit quality.
Similar to optimizing other placement metrics like wirelength and routability~\cite{ripplefpga, utplacef, ripple_2, gplace_3, routing_architecture_aware_placement}, SLL minimization should be an ongoing and holistic process throughout the entire placement.
Further, during the LG and DP, the movement of placeable instances is critical to the potential impact of placement metrics~\cite{utplacef_2, clock_aware_place_Kuo, utplacef2x, clock-aware_ultrascale_fpga_placement, clock_aware_place_Chen}.
This movement may increase the number of SLLs, necessitating positive measures to maintain circuit performance.

Accordingly, our proposed method aims to address three major challenges: 1) simultaneous optimization of various design constraints and objectives, such as wirelength, routability, and clock routing; 2)  capability of adapting to more complex topologies; and 3) holistic optimization of the number of SLLs throughout the placement stage. 
To tackle the above challenges, this paper presents \texttt{LEAPS}, a comprehensive, systematic, and adaptable multi-die FPGA placement algorithm for SLL minimization. Our contribution can be summarized as follows.
\begin{itemize}
	\item We propose a high-performance nested optimization hierarchy for global placement of multi-die FPGA, which aims to reduce wirelength and the number of SLLs, and meanwhile satisfy routability and clock routing constraints.
	\item We introduce an adaptive wirelength-weighting-factor adjusting technique, primarily aimed at balancing the trade-offs between HPWL and SLL counts. 
	This approach is pivotal in achieving a more finely-tuned and optimized placement solution, addressing the wirelength handling challenges in multi-die FPGA design.
	\item We design a flexible method to adapt multi-die FPGAs with arbitrary SLR topologies. It converts SLR indexes into vector representations of instances’ coordinates and uses a soft floor technique, thus enabling a seamless transition from global to local optimization.
	\item We propose a simple but effective optimization technique for SLL minimization at the LG and DP stages.
\end{itemize}
In summary, this paper presents a novel approach to address the challenges in multi-die FPGA placement, specifically
for SLL minimization, while maintaining a focus on other essential design constraints. Our proposed \texttt{LEAPS} framework
demonstrates adaptability to complex topologies and ensures continuous optimization throughout the entire placement process. The experimental results show that our method greatly outperforms the SOTA algorithm~\cite{analytical_place_3d_poisson}.
It achieves significant reductions of 43.08\% and 9.99\% in SLL and HPWL, respectively, while exhibiting a substantial improvement of 34.34$\times$ in runtime.

The rest of the paper is structured as follows. 
\secRef{sec:preliminaries} provides essential background information and formulates
the problem of multi-die FPGA placement addressed by the proposed framework.
\secRef{sec:leaps_overview} presents a comprehensive overview of the \texttt{LEAPS} framework, highlighting its key features and technical innovations. 
\secRef{sec:core_algorithms} delves into the technical details of the core placement algorithms employed in \texttt{LEAPS}. 
\secRef{sec:experimental_results} presents the experimental results, which validate the efficacy and superiority of our approach. 
Finally, \secRef{sec:conclusion} concludes the paper by summarizing the key contributions and highlighting avenues for future research.

\section{Preliminaries}
\label{sec:preliminaries}
In this section, we provide the background and concepts related to the multi-die FPGA placement problem addressed in this paper. 
First, we introduce the multi-die FPGA architecture and its various topologies, as well as the calculation of the number of SLLs and clocking constraints within SLRs.
Then, we also discuss the multi-electrostatic approach used to optimize the placement, emphasizing the advancement of the underlying methods on which our framework depends. 
Finally, we formally state the problem of multi-die FPGA placement, highlighting the key objectives and constraints to be considered in the proposed placement algorithm.

\subsection{Multi-Die FPGA Architecture}
The multi-die FPGA architecture utilizes stacking technology to interconnect multiple FPGA cores, known as SLRs,
via SLLs on an interposer, as depicted in~\figRef{fig:Multi-die FPGA Arch.}\textcolor{blue}{(a)}.
It is worth noting that SLLs have significantly higher delay compared to regular interconnects, which can greatly impact the design’s timing performance and circuit quality.
\figRef{fig:Multi-die FPGA Arch.}\textcolor{blue}{(b)} illustrates that each SLR contains multiple distinct clock regions.
This arrangement facilitates more flexible and efficient clock signal management and routing.

Additionally, each SLR comprises millions of logic gates, including heterogeneous blocks such as look-up tables (LUTs), flip-flops (FFs), digital signal processors (DSPs), random access memories (RAMs), and other intellectual property (IP) blocks. LUTs and FFs are ultimately clustered in configurable logic blocks (CLBs) for placement.
\figRef{fig:Multi-die FPGA Arch.}\textcolor{blue}{(c)}  illustrates the representation of CLBs, which are classified into two types: SLICEL and SLICEM, showing asymmetric compatibility. SLICEL allows LUT blocks to be configured as LUTs, while SLICEM can be configured in one of the following modes: LUT, distributed RAM, or SHIFT, without intermixing LUTs, distributed RAMs, and SHIFTs within a CLB.

\begin{figure}
	\centering
	\includegraphics[width=0.76\columnwidth]{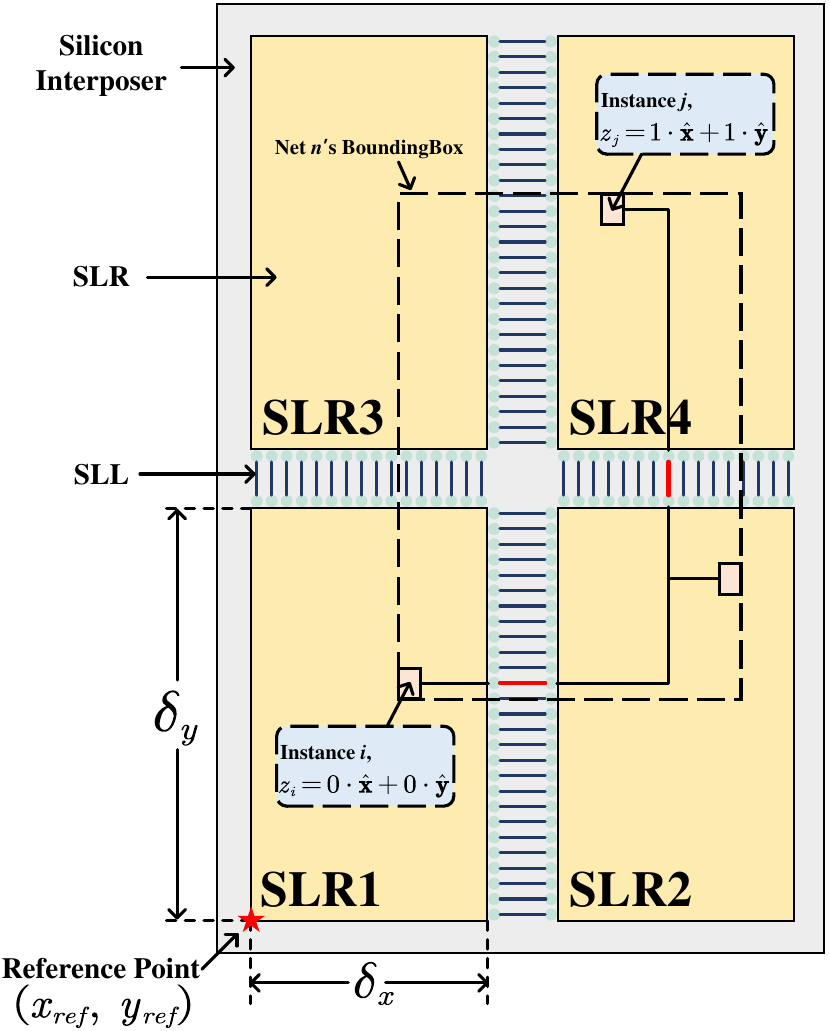}
	\caption{
		Schematic example of a multi-die FPGA featuring a 2$\times$2 SLR topology with an illustrative SLL calculation for a 3-pin net $n$.
	}
	\label{fig:SLL_Calculation_Method.}
\end{figure}

An important aspect of multi-die FPGAs is the arrangement of SLRs, which we refer to as SLR topology, with examples
including $1 \times 4$ (shown as~\figRef{fig:Multi-die FPGA Arch.}\textcolor{blue}{(a)}) and $2\times2$ (shown as \figRef{fig:SLL_Calculation_Method.}) configurations.
An $m \times n$ SLR topology implies an arrangement of $m$ column and $n$ rows of SLRs. 
We present two industrial examples, the Xilinx Alveo U250 FPGA and the Xilinx Alveo U280 FPGA, to illustrate different SLR topologies and some basic configurations.
\begin{itemize}
	\item The Xilinx Alveo U250 FPGA features a $1 \times 4$ SLR topology, with I/O banks and DDR controller IPs located in the middle column, and a Vitis platform region on the right side for communication with the host CPU.
	\item The Xilinx Alveo U280 FPGA, which integrates High-Bandwidth Memory (HBM), has a $2 \times 2$ SLR topology, with I/O banks in the middle columns and a gap region devoid of programmable logic in the center of the chip.
\end{itemize}

Two representative multi-die FPGA architectures are presented above, highlighting the diversity of SLR topologies.
This diversity emphasizes the need for a placement algorithm that can handle different topologies and constraints while ensuring flexibility and efficiency.
However, due to the limitations in academic datasets, our framework design and testing focus on the \emph{Xilinx UltraScale} architecture.

\subsection{SLL Calculation Method}
Calculating the number of SLLs is crucial for the effective placement optimization of multi-die FPGAs. 
The number of SLLs is determined by the number of times a net has to cross between different SLRs.

Given a hypergraph-based placement result, we define the set of placeable instances as $V=\{v_1, v_2, ..., v_n\}$, and the net as $E=\{e_1, e_2, ..., e_n\}$. 
For the multi-die FPGA placement problem, we define the coordinates of instance $v_i$ as $(x_i, y_i, z_i)$, where $x_i$ and $y_i$ represent the physical location of the instance on the layout, and $z_i$ denotes the index of the SLR in which the instance is located. With the above definitions, we can calculate the total number of SLLs as follows:
\begin{equation}
	\label{sll calculation}
    S_{SLL}=\sum_{e \in E} 	f(\{z_i | v_i \in e \}, \mathcal{T}_{SLR}).
\end{equation}
Here, $\{z_i | v_i \in e \}$ denotes the SLR index set of the instances associated with net $n$, and $\mathcal{T}_{SLR}$ denotes the SLR topology. 
The function $f(\cdot)$ denotes the mapping between the specified index set and the SLR topology.
For the $1\times 4$ SLR topology in the previous discussion, this mapping can be found in the existing work~\cite{analytical_place_3d_poisson}.
For complex SLR topologies like $2 \times 2$ or $3 \times 3$, we use a minimum spanning tree (MST) for mapping, which is efficient due to the typically small size of these SLR topologies (with rows and columns less than 5). 
A mapping table in our function further improves computational speed.
Notably, while our method can adapt to any SLR topologies, choosing the best one involves balancing the benefits of multi-die architectures against practical constraints. 
These include clock region division, timing closure, and manufacturing factors related to cost and limitations.

To determine the SLR index $z_i$ for each instance, we use the Manhattan metric considering both $x_i$ and $y_i$.
We define the distance thresholds $\delta_x$ and $\delta_y$ as the width and height of each SLR, respectively. 
The width and height of an SLR are computed by dividing the total width and height of the FPGA's layout by the number of columns and rows in the SLR topology. 
In our method, the reference point $(x_{ref}, y_{ref})$ is set to the bottom-left corner of the layout $(0, 0)$.
With these definitions, the SLR index $z_i$ can be calculated as follows:
\begin{equation}
	\label{eq:general_z_calculation}
	\boldsymbol{z}_i = \lfloor \frac{|x_i - x_{ref}|}{\delta_x} \rfloor \cdot \hat{\mathbf{x}} + \lfloor \frac{|y_i - y_{ref}|}{\delta_y} \rfloor \cdot \hat{\mathbf{y}},
\end{equation}
where $\hat{\mathbf{x}}$ and $\hat{\mathbf{y}}$ represent the unit vectors in the $x$ and $y$ dimensions, respectively.
Note that the SLR index $\boldsymbol{z}_i$ is represented as a two-dimensional vector. 
This vector representation will be consistently used in subsequent sections for clarity and uniformity.

The above method for computing SLLs is an improved version of the method in~\cite{analytical_place_3d_poisson}.
It’s proposed to accommodate SLR topologies with multiple rows and columns, such as $2\times2$ SLR topology. 
The motivation behind this improved method is the indispensable role that precise quantization of the number of SLLs for efficient optimization. 
This precise quantification allows the placer to adjust the arrangement of logic instances, thus facilitating the minimization of the total number of SLLs.

\subsection{Clocking Constraints in SLRs}
Clocking constraints are crucial for both performance optimization and timing closure in a multi-die FPGA design flow. 
The target device has rectangular-shaped clock regions (CRs) arranged in a $5 \times 8$ grid, each consisting of columns of site resources. 
The CRs can be further subdivided horizontally into upper and lower half-columns (HCs), with a maximum of 12 clock nets per HC and a maximum of 24 clock nets per CR.
These constraints are referred to as the half-column constraint and the clock region constraint, respectively.

To mathematically model these constraints, we first define $C_{k}^{z}$ as the set of blocks connected to clock $k$ on SLR $z$. 
We then establish the $x\left(y\right)$ coordinates of the right, left (top, bottom) boundaries of clock region $o$ on SLR $z$, denoted by $r_{o}^{z}$, $l_{o}^{z}\left(u_{o}^{z}, d_{o}^{z}\right)$, respectively. 
Accordingly, we can calculate the horizontal and vertical clocking resource usages for clock $k$ in clock region $o$ on SLR $z$ as:
\begin{equation}
	\begin{aligned}
		& H(k, o, z)=\min \left\{\max \left\{x_i \mid i \in C_k^z\right\}, r_o^z\right\} \\
		& \qquad\qquad -\max \left\{\min \left\{x_i \mid i \in C_k^z\right\}, l_o^z\right\}, \\
		& V(k, o, z)=\min \left\{\max \left\{y_i \mid i \in C_k^z\right\}, u_o^z\right\} \\
		& \qquad\qquad -\max \left\{\min \left\{y_i \mid i \in C_k^z\right\}, d_o^z\right\} .
	\end{aligned}
\end{equation}

The total clock usage $P(k, o, z)$ for clock $k$ in clock region $o$ on SLR $z$ can be computed as:
\begin{equation}
	P(k, o, z)= \begin{cases}1, & \text { if } H(k, o, z)>0 \text { and } V(k, o, z)>0; \\ 0, & \text { otherwise. }\end{cases}
\end{equation}

By assuming that clock region $o$ on SLR $z$ is covered by at most $M_{o,z}$ clock net bounding boxes, we can define the clocking constraints of a multi-die FPGA as:
\begin{equation}
	\sum_k P(k, o, z) \leq M_{o, z}, \quad \forall o, z.
\end{equation}

With the above definitions, we effectively model the clocking constraints in multi-die FPGA architectures. 
This model allows a more precise depiction of the clocking resources and their existing constraints within the FPGA device. 
As a result, it will facilitate the development process of more advanced multi-die FPGA placement algorithms, making the overall efficiency higher.

\subsection{Multi-Electrostatic FPGA Placement}
State-of-the-art placement algorithms\cite{elfPlace_Meng, multi-electrostatic_placement, openparf, dreamplace, dreamplacefpga}, grounded in electrostatics, conceptualize each instance as a positive charge within an electrostatic system. 
This approach was originally introduced in ASIC placement to mitigate density overflow problem in the placement, and leverages the physical principle of balanced charge distribution leading to low potential energy in electrostatic systems.
\cite{elfPlace_Meng} expanded this method to include multiple electrostatic fields, thereby facilitating the management of diverse resource types in FPGA placement, such as LUTs, FFs, DSPs, and BRAMs. 
Building upon these advancements, recent work~\cite{multi-electrostatic_placement, openparf} has further refined the multi-electrostatic approach by incorporating considerations of SLICEL-SLICEM heterogeneity and multiple constraints, including timing, clock routing, and carry chain alignment.
This innovative algorithm takes the quality and efficiency of FPGA placement a significant step forward, surpassing its predecessors. 
It seeks to minimize the total potential energy of multiple fields, effectively reducing density overflow. 
Given this capacity for adept resource distribution management across multiple dies and their respective clocking domains, the multi-electrostatic approach is especially well-suited to multi-die FPGA placement. 
The primary objective of this approach is to optimize placement by achieving a balanced resource distribution in the layout. 
This problem can be mathematically formulated as follows:
\begin{equation}
	\min _{\boldsymbol{x}, \boldsymbol{y}} \widetilde{W}(\boldsymbol{x}, \boldsymbol{y}) \quad \text { s.t. } \Phi_s(\boldsymbol{x}, \boldsymbol{y})=0, \forall s \in S
\end{equation}
Here, $\boldsymbol{x}, \boldsymbol{y}$ represent instances' location, $\widetilde{W}(\cdot)$ denotes the wirelength objective, $S$ is the field type set, and $\Phi_s(\cdot)$ signifies the electric potential energy of the field for field type $s \in S$. We formally constrain the target energy $\Phi_s(\boldsymbol{x}, \boldsymbol{y})$ to $0$, as the energy is typically nonnegative. 
The constraints can be relaxed to the objective and solved using the gradient descent method. 
In practice, optimization is ceased when the energy reaches a sufficiently low level, or equivalently when the density overflow reaches an acceptable threshold.

\subsection{Problem Statement for Multi-Die FPGA Placement}
In this paper, we aim to solve the problem of multi-die FPGA placement. 
The optimization objective is to minimize the half-perimeter wirelength (HPWL) and the number of SLLs crossing SLRs while satisfying various architectural constraints.
Formally, the problem can be formulated as follows:
\newtheorem{problem}{Problem}
\begin{problem}[Multi-Die FPGA Placement]
	Given a circuit netlist $\mathcal{N}$, a placement region $\mathcal{R}$, and architecture constraints $\mathcal{C}$, determine the optimal legal position $(x_i, y_i)$ of each logic block $i$ on a SLR to minimize the HPWL $W_\mathcal{H}$ and the number of SLLs $W_\mathcal{S}$ with a weighting factor $\psi$, such that all architecture constraints in $\mathcal{C}$ are satisfied. Mathematically, the problem can be formulated as:
	\begin{equation}
		\begin{aligned}
			\min_{\boldsymbol{x}, \boldsymbol{y}} \hspace{0.2em}&W_\mathcal{H}(\boldsymbol{x}, \boldsymbol{y}) + \psi W_\mathcal{S}(\boldsymbol{x}, \boldsymbol{y}) \\
			\text{s.t. } &\boldsymbol{x}, \boldsymbol{y} \in \mathcal{R}, \\
	        &\nexists\hspace{0.2em} i, j \in \mathcal{N}\text{ with }i \neq j \text{ such that }\mathcal{O}\mkern-2mu\text{verlap}(i, j) > 0,\\
			&\text{all architecture constraints in }\mathcal{C}\text{ are satisfied}.
		\end{aligned}
	\end{equation}
\end{problem}

\section{LEAPS Overview}
\label{sec:leaps_overview}
\begin{figure}
	\centering
	\includegraphics[width=\columnwidth]{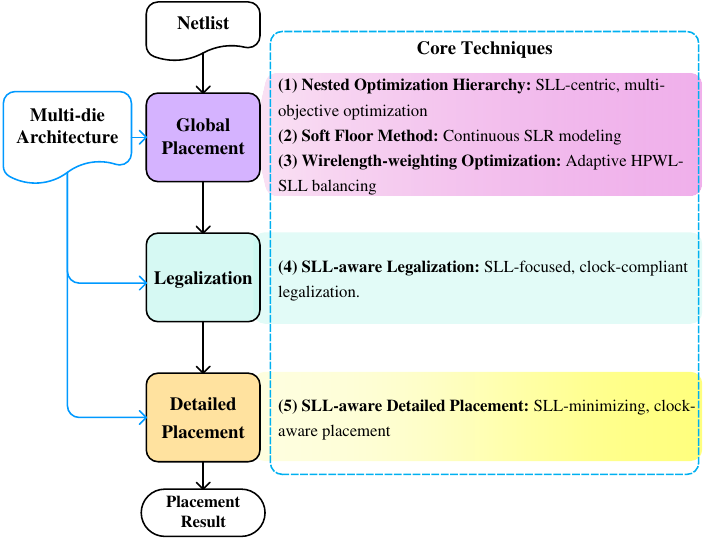}
	\caption{
		Core Techniques in the proposed \texttt{LEAPS}:
		(1) \textit{Nested Optimization Hierarchy}: Enhances \cite{multi-electrostatic_placement, openparf} for multi-objective optimization, focusing on SLL minimization. See \secRef{subsubsec.:Nested Optimization Hierarchy}.
		(2) \textit{Soft Floor Method}: Transforms discrete SLR coordinates into continuous models, optimizing wirelength and SLR constraints. Refer to \secRef{sec:SoftFloorMethod}.
		(3) \textit{Wirelength-weighting Optimization}: Dynamically adjusts HPWL and SLL trade-offs for improved FPGA placement. Details in \secRef{subsec.:Adaptive Wirelength-Weighting-Factor Adjusting}.
		(4) \textit{SLL-aware Legalization}: Adapts \cite{direct_legalize} to prioritize SLL reduction with concurrent clock constraint management. Further information in \secRef{subsec.:Clock- and SLL-aware Legalization & Detailed Placement}.
		(5) \textit{SLL-aware Detailed Placement}: Builds on \cite{utplacef}, focusing on SLL minimization, clock-awareness, and wirelength optimization. See \secRef{subsec.:Clock- and SLL-aware Legalization & Detailed Placement}.
	}
	\label{fig:core_techniques}
\end{figure}

Building upon the foundational contributions of \cite{openparf, multi-electrostatic_placement}, we introduce the proposed \texttt{LEAPS} for multi-die FPGA placement, consisting of three main stages: global placement, legalization, and detailed placement.
A summary of the core techniques employed in the proposed \texttt{LEAPS} is provided in~\figRef{fig:core_techniques}.
Furthermore, to ensure a coherent explanation of key terminologies used throughout this paper, a detailed glossary is presented in~\tabRef{tab:terminology}.

\begin{table*}[t]
	\centering
	\caption{Glossary Table of Key Terminology in the proposed \texttt{LEAPS}}
	\resizebox{\textwidth}{!} {
		\begin{tabular}{|m{0.2\textwidth}|m{0.4\textwidth}|m{0.4\textwidth}|}
			\hline
			\multicolumn{1}{|c}{\textbf{Term}} & \multicolumn{1}{|c|}{\textbf{Definition}} & \multicolumn{1}{c|}{\textbf{Motivations}} \\
			\hline
			Augmented Lagrangian Method (ALM) & Transforms constrained problems into unconstrained ones by adding equality constraints and a quadratic penalty term. & Simplifies \eqRef{eq:constrained_problem}'s complex constrained problem into \eqRef{eq:opt_alm}'s unconstrained problem. \\
			\hline
			Clock Penalty Multiplier $\eta$ & Dynamically penalizes clock routing violations in the optimization process. & Ensures clock routing constraints are considered with other design objectives. \\
			\hline
			Density Penalty $D_s$ & Penalizes denser areas in the FPGA placement to even out logic element distribution. & Prevents hotspots and ensures routability in FPGA placement. \\
			\hline
			Wirelength-Weighting-Factor $\Psi$ & Balances the minimization of HPWL and the reduction of SLL counts. & Manages HPWL and SLL trade-offs in multi-die FPGA design. \\
			\hline
			Instance-to-Clock-Region Mapping Generation& Assigns instances to clock regions to minimize SLLs. & Optimizes clock network efficiency and FPGA performance. \\
			\hline
		\end{tabular}
	}

	\label{tab:terminology}
\end{table*}

The proposed \texttt{LEAPS} considers a set of field types denoted as $S = \{ \text{LUTL, LUTM-AL, FF, CARRY, DSP, BRAM} \}$. 
Notably, the LUTL and LUTM-AL field types were introduced in~\cite{multi-electrostatic_placement}. 
The LUTL field type represents the LUT resources provided by both SLICEM and SLICEL, while the LUTM-AL field type models the additional logic resources offered solely by SLICEM and not by SLICEL.

\subsection{Global Placement}
Global placement acts as the backbone of the entire placement algorithm, harmonizing multiple design objectives while satisfying complex constraints. 
\subsubsection{Problem Definition}
Considering wirelength minimization objective, clock constraints, and carry chain alignment feasibility, we present the multi-die global placement problem as a constrained minimization problem:
\begin{subequations}
\label{eq:constrained_problem}
\begin{align}
	\min_{\boldsymbol{x}, \boldsymbol{y}}&\quad \widetilde{W}_{\psi}(\boldsymbol{x}, \boldsymbol{y}), \\
	\text {s.t.}&\quad \Phi_s\left(\boldsymbol{x}, \boldsymbol{y} ; \mathcal{A}^s\right)=0, \quad \forall s \in S, \\
	&\quad \varGamma(\boldsymbol{x}, \boldsymbol{y})=0, \\
	&\quad \textit{Carry chain alignment constraint.} 
\end{align}
\end{subequations}
Here, $\widetilde{W}_{\psi}(\boldsymbol{x}, \boldsymbol{y})$ denotes the total wirelength, which accounts for both the HPWL and SLL counts;
$\mathcal{A}^s$ represents all the instance areas in field $s$; 
and $\varGamma(\cdot)$ signifies the clock penalty term. 
For brevity, we simplify $\Phi_s\left(\boldsymbol{x},\boldsymbol{y};\mathcal{A}\right)$ to $\Phi_s$ for all $ s \in S$. 
The potential energy vector, with elements $\Phi_s\left(\forall s \in S\right)$, is denoted by $\boldsymbol{\Phi}$ in subsequent discussions.

\subsubsection{Problem Reformulation with ALM}
To facilitate solving the original problem~(\ref{eq:constrained_problem}), we employ the \textit{augmented Lagrangian method (ALM)}~\cite{alm} to formulate an unconstrained subproblem:
\begin{subequations}
	\begin{align}
		\min_{\boldsymbol{x}, \boldsymbol{y}} \quad \mathcal{L}(\boldsymbol{x}, \boldsymbol{y}; \boldsymbol{\lambda}, \psi, \boldsymbol{\mathcal{A}},\eta) & =
		\widetilde{W}_{\psi}(\boldsymbol{x}, \boldsymbol{y})  + \sum\limits_{s \in S} \lambda_s \mathcal{D}_s \nonumber \\
		& + \eta \varGamma(\boldsymbol{x}, \boldsymbol{y}), \\
		\mathcal{D}_s &= \Phi_s + \frac{1}{2}\mathcal{W}_s \Phi_s^2,\quad \forall s \in S.
	\end{align}
	\label{eq:opt_alm}
\end{subequations}
Here, the density multiplier vector is $\boldsymbol{\lambda} \in \mathbb{R}^{\left|S\right|}$, and the clock penalty multiplier is $\eta \in \mathbb{R}$. 
The density-weighting coefficient vector $\boldsymbol{\mathcal{W}} \in \mathbb{R}^{\left|S\right|}$ is employed to balance the first-order and second-order terms for density penalty. 
We adopt the setup for $\boldsymbol{\lambda}$ and $\boldsymbol{\mathcal{W}}$ from~\cite{elfPlace_Meng}. 
\subsubsection{Nested Optimization Hierarchy}
\label{subsubsec.:Nested Optimization Hierarchy}
To handle multiple constraints, we solve the problem~(\ref{eq:opt_alm}) using the ALM in a nested manner:
\begin{subequations}
	\begin{align}
		\textrm{Clock Opt.: } \mathcal{L}_1 & = \max_{\eta} \mathcal{L}_2 (\eta), \label{eq:clock opt.}\\
		\textrm{Routability Opt.: } \mathcal{L}_2(\eta) & = \max_{\boldsymbol{\mathcal{A}}} \mathcal{L}_3 (\boldsymbol{\mathcal{A}}, \eta),  \label{eq:routability opt.}\\
		\textrm{WLW Opt.: } \mathcal{L}_3(\boldsymbol{\mathcal{A}}, \eta) & = \max_{\psi} \mathcal{L}_4(\psi, \boldsymbol{\mathcal{A}},\eta) \label{eq:wlw opt.}\\
		\textrm{WL Opt.: } \mathcal{L}_4(\psi, \boldsymbol{\mathcal{A}}, \eta) & = \max_{\boldsymbol{\lambda}} \mathcal{L}_5 ( \boldsymbol{\lambda}, \psi,  \boldsymbol{\mathcal{A}}, \eta),\label{eq:wl opt.} \\
		\textrm{Subproblem: }\mathcal{L}_5(\boldsymbol{\lambda}, \psi, \boldsymbol{\mathcal{A}}, \eta) & = \min_{\boldsymbol{x}, \boldsymbol{y}} \mathcal{L}(\boldsymbol{x}, \boldsymbol{y}; \boldsymbol{\lambda}, \psi, \boldsymbol{\mathcal{A}},\eta) \label{eq:subproblem}.
	\end{align}
	\label{eq:nested_opt}
\end{subequations}
Here, $\mathcal{L}_{5}$ denotes \eqRef{eq:opt_alm}; ``Opt.'' stands for ``Optimization'', while ``WL'' and ``WLW'' are abbreviations for ``Wirelength'' and ``Wirelength-weighting'', respectively.

\begin{figure}[t]
	\centering
	\includegraphics[width=.9\columnwidth]{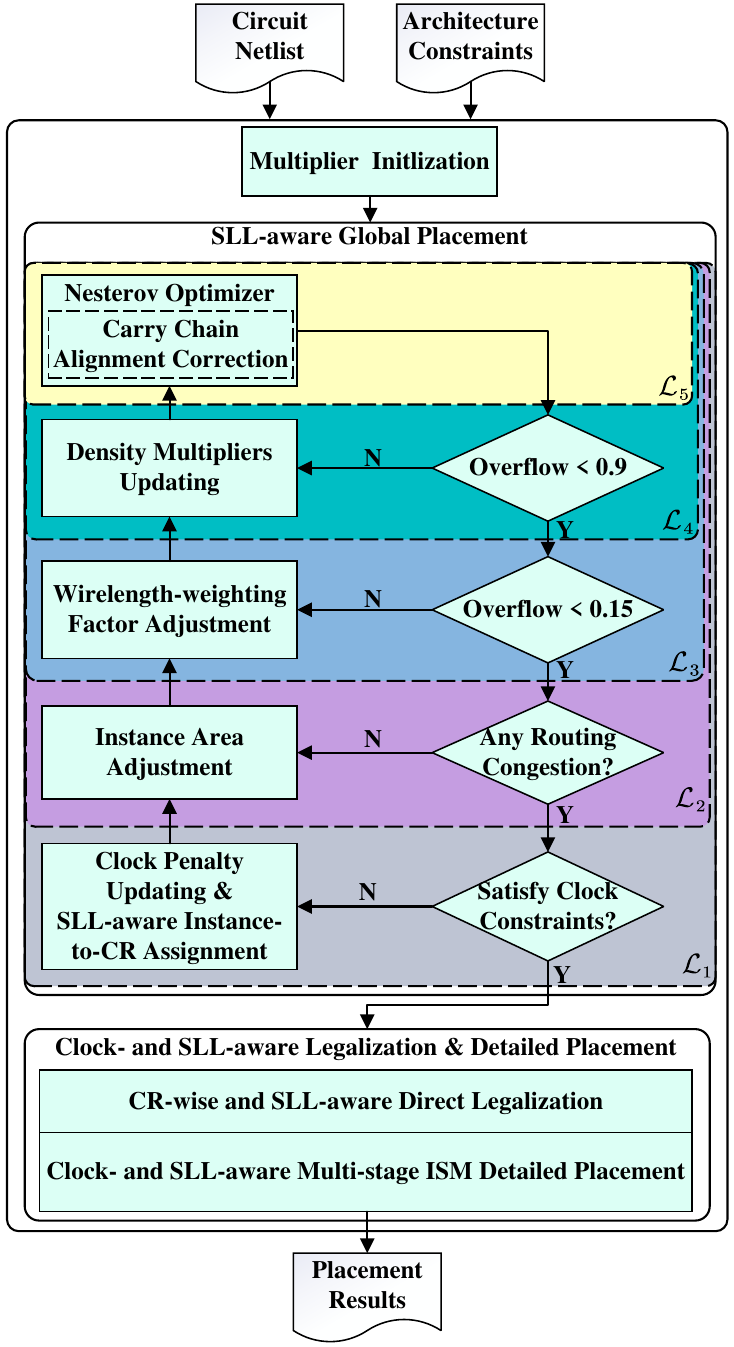}
	\caption{Overview of the proposed \texttt{LEAPS} framework.
		The framework continuously optimizes the number of SLLs while handling other design objectives during the global placement, legalization, and detailed placement stages.
		The global placement employs a nested optimization technique to progressively converge and optimize each design objective.
		Subsequent legalization and detailed placement consider SLL minimization and clock routing constraints while refining the initial placement.
		Note that ``Instance Area Adjustment'' and ``Carry Chain Alignment Correction''are referenced in~\cite{elfPlace_Meng}~and~\cite{multi-electrostatic_placement}, respectively, and are not repeated in this paper. The rest of the contents are described in this work.}
	\label{fig:The Proposed Framework}
\end{figure}

Within this nested structure, each term, ranging from $\mathcal{L}_1$ to $\mathcal{L}_5$, addresses a unique aspect of the placement challenge. 
Each term passes its variables to the subsequent subproblem, considering them as fixed hyperparameters. 
This systematic approach is vividly portrayed in the global placement phase as shown in \figRef{fig:The Proposed Framework}.
We highlight two vital aspects:
\paragraph{Effective Range for SLL Optimization}
The SLL minimization is integrated into the optimization objective only when the density overflow lies between 0.15 and 0.9. 
Within these bounds, $\mathcal{L}_4$ becomes operative. 
When the density overflow drops below 0.15, the algorithm adjusts the instance area to mitigate routing congestion.

\paragraph{Distinct Roles of Each Optimizer}
To illustrate the overall approach, we consider the example of optimizer $\mathcal{L}_5$, particularly in scenarios where density overflow exceeds 0.9. 
Upon reaching its optimal state, $\mathcal{L}_5$ maintains the parameters $\boldsymbol{\lambda}$, $\psi$, $\boldsymbol{\mathcal{A}}$, and $\eta$ as fixed values.
Subsequently, the $\mathcal{L}_4$ optimizer amplifies the density term by incrementing $\boldsymbol{\lambda}$, progressively satisfying the density constraints.
This approach is similarly employed for other optimizers. 
Each optimizer has a specific role:
\begin{itemize}
\item $\mathcal{L}_{1}$ analytically mitigates clock violations, with its termination criteria based on the fulfillment of clock constraints. 
\item $\mathcal{L}_{2}$ enhances routability via an area inflation-based technique, with its termination criteria determined by routing congestion estimation and pin density.
\item $\mathcal{L}_{3}$ accounts for the growth of SLLs during the iterative process to balance the trade-off between the HPWL and SLL counts in the total wirelength objective.
\item $\mathcal{L}_{4}$ tackles the core wirelength-driven placement problem, with its termination criteria determined by the density overflow of all instances. 
\end{itemize}
In the end, $\mathcal{L}_{5}$ is consistently solved with a pre-defined number of iterations, such as one iteration in our experiments.

\subsection{Legalization \& Detailed Placement}
Legalization (LG) and detailed placement (DP) serve as the refining stages of the placement, fine-tuning the initial solution and ensuring compliance with design constraints and objectives. 
Given the multi-die FPGA architecture, both LG and DP need to achieve the aim of minimizing SLLs under various design constraints.
To achieve this, we adapt our LG and DP to specifically target the reduction of SLLs while maintaining other design objectives.

Our LG approach, inspired by the Direct Legalize (DL) algorithm~\cite{direct_legalize}, employs a binary optimization strategy to map instances to specific clock regions. 
This method has been adapted and expanded from its original single-die application to multi-die FPGAs, incorporating a refined cost function that includes SLL optimization. 
This enhancement makes it better suited to address the unique challenges presented in multi-die scenarios, going beyond the traditional DL method by considering factors like HPWL, SLL counts, and packing metrics. 
This approach not only ensures a pronounced reduction in the number of SLLs but also maintains essential clock feasibility constraints.

Complementing this, our DP method builds on the multi-stage independent set matching (ISM) technique as presented in~\cite{utplacef, utplacef_2}. 
It emphasizes the reduction of the number of SLLs, similar to the legalization process, and employs clock-awareness to the ISM method, thereby refining placement for improved wirelength and routability. 
This refinement ensures the viability of the clock network and simultaneously optimizes SLL counts, a crucial aspect frequently neglected in conventional single-die FPGA placement strategies.

These approaches, building upon the foundational frameworks of~\cite{utplacef, utplacef_2, direct_legalize}, introduce novel enhancements specifically designed for the complexities of multi-die architectures. 
These advancements are essential to meet the evolving demands of modern FPGA design.

\subsection{Comparative Features of LEAPS and Other Placers}
\label{sec:Comparative Features of LEAPS and Other Placers}
\begin{table*}[ht]
	
	\caption{Features of the published state-of-the-art FPGA placers}
	\label{tab:FPGAPlacers}
	\resizebox{1.0\textwidth}{!}{
		\begin{tabular}{|cc|cccccccccc|}
			\hline
			\multicolumn{2}{|c|}{\textbf{Placer}}
			& \begin{tabular}[c]{@{}c@{}}\texttt{UTPlaceF}\\ \cite{utplacef}\end{tabular}
			& \begin{tabular}[c]{@{}c@{}}\texttt{GPlace 3.0}\\ \cite{gplace_3}\end{tabular}
			& \begin{tabular}[c]{@{}c@{}}\texttt{elfPlace}\\ \cite{elfPlace_Meng}\end{tabular}
			& \begin{tabular}[c]{@{}c@{}}\texttt{DREAMPlaceFPGA}\\ \cite{dreamplace}\end{tabular}
			& \begin{tabular}[c]{@{}c@{}}\texttt{AMF-Placer}\\ \cite{amfplacer}\end{tabular}
			& \begin{tabular}[c]{@{}c@{}}\texttt{UTPlaceF}\\ \texttt{{2.0}}\&\texttt{2.X} \cite{utplacef_2, utplacef2x}\end{tabular}
			& \begin{tabular}[c]{@{}c@{}}\texttt{ICCAD'17}\\ \cite{clock_aware_place_Kuo}\end{tabular}
			& \begin{tabular}[c]{@{}c@{}}\texttt{ICCAD'19}\\ \cite{analytical_place_3d_poisson}\end{tabular}
			& \begin{tabular}[c]{@{}c@{}}\texttt{OpenPARF}\\ \cite{openparf} \end{tabular}
			& \begin{tabular}[c]{@{}c@{}}Ours \end{tabular}         \\ \hline
			\multicolumn{2}{|c|}{Clock Constraints}                                                                                                                                         & $\times$     & $\times$     & $\times$     & $\times$     & $\times$     & $\checkmark$                                              & $\checkmark$                                                     & $\checkmark$                                                & $\checkmark$ & $\checkmark$ 
			\\ \hline
			\multicolumn{2}{|c|}{Multi-die Support}                                                                                                                                       & $\times$     & $\times$     & $\times$     & $\times$     & $\times$ & $\times$                                             & $\times$                                                         & $\checkmark$                                                    & $\times$     & $\checkmark$
			\\ \hline
			\multicolumn{2}{|c|}{GPU-Acceleration}                                                                                                                                          & $\times$     & $\times$     & $\checkmark$     & $\checkmark$ & $\times$     & $\times$                                             & $\times$                                                         & $\times$                                                    & $\checkmark$     & $\checkmark$     
			\\ \hline
			\multicolumn{2}{|c|}{Algorithm Category}                                                                                                                                        & Quadratic    & Quadratic    & Nonlinear    & Nonlinear     & Quadratic    
			& Quadratic                                   & Quadratic                                                        & Nonlinear                                                   & Nonlinear    & Nonlinear
			\\ \hline
	\end{tabular}}
\end{table*}
\tabRef{tab:FPGAPlacers} summarizes the characteristics of the published SOTA FPGA placers.
They mainly resort to quadratic programming-based approaches~\cite{gplace_3, utplacef, utplacef_2, utplacef2x, amfplacer, clock_aware_place_Kuo} and nonlinear optimization-based approaches~\cite{elfPlace_Meng, dreamplacefpga, analytical_place_3d_poisson, openparf} for the best trade-off between quality and efficiency.
To provide a comprehensive assessment, we categorize our evaluation based on three pivotal features:
\begin{itemize}
	\item \textit{Handling Clock Constraints}: An important issue in FPGA placement is to effectively address clock constraints to optimize performance. Among these SOTA placers, \texttt{UTPlaceF 2.0\&2.X}~\cite{utplacef_2, utplacef2x}, \texttt{ICCAD'17}~\cite{clock_aware_place_Kuo}, \texttt{ICCAD'19}~\cite{analytical_place_3d_poisson}, \texttt{OpenPARF}~\cite{openparf}, and the proposed \texttt{LEAPS} exhibit proficiency in this domain.
	\item \textit{Supporting Multi-Die Architecture}: The capability to facilitate designs across multiple dies is only present in \texttt{ICCAD'19}\cite{analytical_place_3d_poisson} and the proposed \texttt{LEAPS}. 
	This gives them an advantage in modern FPGA designs, where multi-die configurations are sought for enhanced performance and modularity.
	\item \textit{Leveraging GPU acceleration}: Speeding up the placement process is crucial in FPGA design. Among the placers, only \texttt{elfPlace}~\cite{elfPlace_Meng}, \texttt{DREAMPlaceFPGA}~\cite{dreamplacefpga}, \texttt{OpenPARF}~\cite{openparf}, and the proposed \texttt{LEAPS} capitalize on GPU acceleration, making them ideal for rapid design iterations.
\end{itemize}

Conclusively, \texttt{LEAPS} demonstrates a robust feature set aligning with modern FPGA design demands, including clock constraints, multi-die support, and GPU acceleration, making it stand out in the FPGA placement landscape.

\section{Core Placement Algorithms}
\label{sec:core_algorithms}
In this section, we will explicate the core algorithms of the proposed framework.

\subsection{Wirelength Objective Handling}
The wirelength is the most fundamental objective in placement algorithms. 
In traditional FPGA placement algorithms, wirelength is typically measured in the $\boldsymbol{x}$ and $\boldsymbol{y}$ dimensions. 
However, considering the multi-die FPGA architectures in this work, it is necessary to minimize the number
of SLLs by incorporating the $\boldsymbol{z}$ dimension, which represents the SLR index of instances dominated by  $\boldsymbol{x}$ and $\boldsymbol{y}$.

\subsubsection{Wirelength Objective Formulation}
The wirelength objective is formulated as:
\begin{equation}
	\begin{aligned}
		W_{\psi}\left(\boldsymbol{x},\boldsymbol{y}\right) &= 
		W_\mathcal{H}\left(\boldsymbol{x}, \boldsymbol{y}\right) + \psi \cdot W_\mathcal{S}\left(\boldsymbol{x}, \boldsymbol{y}\right) \\
		&= \sum_{e \in E} \left(
		\max_{i, j \in e} \left|x_i - x_j\right|
		+ \max_{i, j \in e} \left|y_i - y_j\right| \right.\\
		&\qquad\left.+ \psi \max_{i, j \in e} \rVert \boldsymbol{z}_i - \boldsymbol{z}_j \rVert_1
		\right)
	\end{aligned}
	\label{eq:origal_wirelength_obj}
\end{equation}
Here, $\boldsymbol{x}$ and $\boldsymbol{y}$ denote the locations of the instances in a layout, while $\boldsymbol{z}$ represents the SLR index of instances dominated by $\boldsymbol{x}$ and $\boldsymbol{y}$.
The term $\rVert \cdot \rVert_1$ denotes the L1 norm, applicable here as the SLR indexes $\boldsymbol{z}_i$ and $\boldsymbol{z}_j$ are two-dimensional vectors. 
The weighting factor $\psi$ adjusts the weighting of the SLL term $W_\mathcal{S}(\cdot)$ in the wirelength objective function $W_{\psi}(\cdot)$.

\subsubsection{Smooth and Differentiable Wirelength Model}
To enable the utilization of gradient-based optimization methods, we adopt a smooth and differentiable wirelength model using the \textit{weighted-average (WA) approach}~\cite{tsv-aware_tcad'13} for the max term. 
Specifically, the wirelength model for the $z$-dimension is defined as:
\begin{equation}
	\begin{aligned}
		\widetilde{W}_{e_{z}}\left(z\right) = 
		\frac{\sum_{i \in e} \rVert \boldsymbol{z}_i \rVert_1 \exp(\rVert \boldsymbol{z}_i \rVert_1 / \gamma_{\mathcal{S}})}{\sum_{i \in e} \exp(\rVert \boldsymbol{z}_i \rVert_1 / \gamma_{\mathcal{S}})} - \\
		\frac{\sum_{i \in e} \rVert \boldsymbol{z}_i \rVert_1 \exp(-\rVert \boldsymbol{z}_i \rVert_1 / \gamma_{\mathcal{S})}}{\sum_{i \in e} \exp(-\rVert \boldsymbol{z}_i \rVert_1 / \gamma_{\mathcal{S}})},
	\end{aligned}
\end{equation}
Here, $\gamma_{\mathcal{S}} > 0$ is a parameter controlling the accuracy of the approximation. As $\gamma_{\mathcal{S}}$ increases, the approximation becomes more accurate, but the objective function becomes less smooth. 
In this work, we utilize the parameter $\gamma_{\mathcal{S}}$ to estimate the wirelength in the $z$-direction, while also introducing a parameter $\gamma_{\mathcal{H}}$ for the wirelength approximation in the $x$ and $y$ directions.
Employing this smooth approximation allows the wirelength model to be differentiable, facilitating the use of gradient-based optimization methods.

By substituting the smooth approximations for the max function in $x$, $y$, and $z$ directions into the original wirelength objective shown as \eqRef{eq:origal_wirelength_obj}, we obtain a smooth and differentiable wirelength objective:
\begin{equation}
	\label{eq:diff_wl_obj}
	\begin{aligned}
	 \widetilde{W}_\psi(\boldsymbol{x}, \boldsymbol{y}) 
	 &= \widetilde{W}_\mathcal{H}\left(\boldsymbol{x}, \boldsymbol{y}\right) + \psi\cdot \widetilde{W}_\mathcal{S} \left(\boldsymbol{x}, \boldsymbol{y}\right) \\
	 &= \sum_{e\in E} \left(
	 \widetilde{W}_{e_{x}}\left(x\right) + \widetilde{W}_{e_{y}}\left(y\right) + \widetilde{W}_{e_{z}}\left(z\right)
	 \right),	
	\end{aligned}	
\end{equation}

This wirelength formulation greatly expands the utility of gradient-based optimization methods within our placement algorithm. 
By minimizing the wirelength objective, the algorithm aims to achieve improved placement results in terms of wirelength while considering constraints related to SLLs.

\subsubsection{Soft Floor Method for Discrete Coordinates $\boldsymbol{z}$}
\label{sec:SoftFloorMethod}
\begin{figure*}
	\centering
	\includegraphics[width=\textwidth]{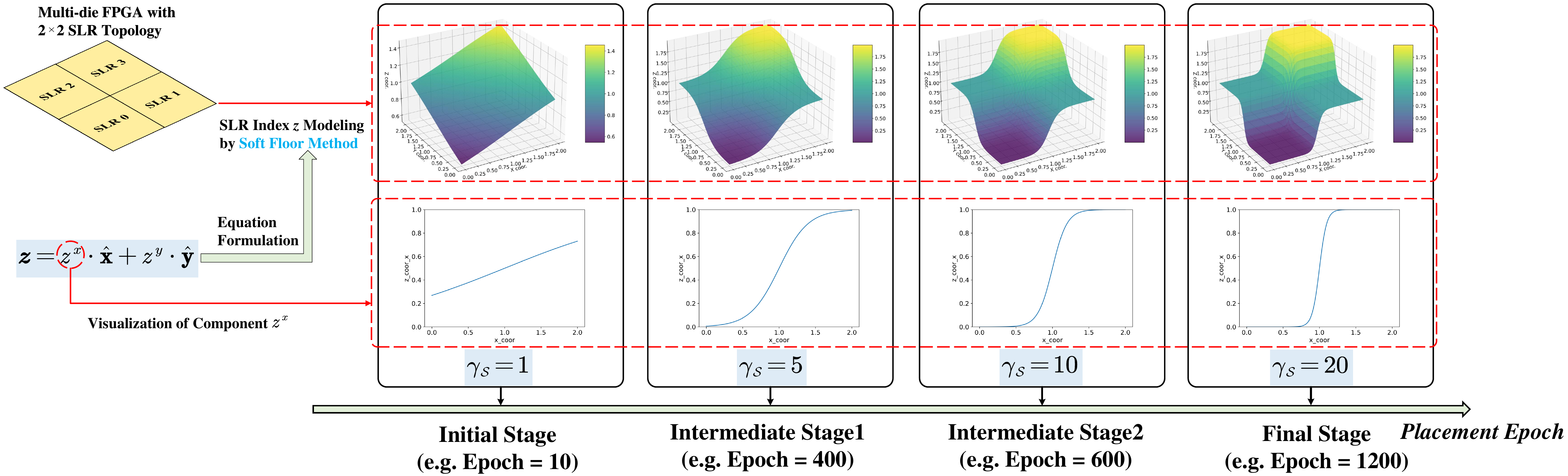}
	\caption{
		Visualization of the soft floor method applied to a multi-die FPGA with a $2\times2$ SLR topology: Demonstrating variations with different $\gamma_{\mathcal{S}}$ values.		
	}
	\label{fig:soft_floor_example}
\end{figure*}
The discrete nature of $\boldsymbol{z}$ presents challenges when optimizing the wirelength objective, as discrete variables can impede the convergence of the optimization algorithm. 
To overcome this issue, we aim to transform the discrete $\boldsymbol{z}$ into a continuous and smooth variable.

Specifically, we propose a soft floor method that enables the smoothing and continuous representation of $\boldsymbol{z}$. 
This approach utilizes a sigmoid-like function, defined as follows:
\begin{equation}
	\sigma(x) = \frac{1}{1+exp(-\gamma_\mathcal{S} \cdot x)}
\end{equation}
Here, $exp(\cdot)$ denotes the exponential function, while $\gamma_{\mathcal{S}}$ is an adaptive parameter used in the wirelength objective for the $z$-dimension, as illustrated in \eqRef{eq:diff_wl_obj}. 
By employing the sigmoid function $\sigma(\cdot)$, a continuous and smooth transformation of $\boldsymbol{z}_{i}$ can be formulated as:
\begin{equation}
	\begin{aligned}
		\boldsymbol{z}_i &= z_{i}^{x} \cdot \hat{\mathbf{x}} + z_{i}^{y} \cdot \hat{\mathbf{y}} \\  
						 &= \sum_{k=0}^{k=\left|z^x\right|-1} \sigma(\frac{x_i}{\sigma_x} -k) \cdot \hat{\mathbf{x}} + 
						 	\sum_{k=0}^{k=\left|z^y\right|-1} \sigma(\frac{y_i}{\sigma_y} -k) \cdot \hat{\mathbf{y}}
	\end{aligned},
\label{eq:soft_floor_trans}
\end{equation}
\eqRef{eq:soft_floor_trans} illustrates how the discrete vector $\boldsymbol{z}$ is transformed into a continuous, smoothly varying two-dimensional vector. 
This transformation is related to the normalized coordinates of the instances $\left(\frac{x_i}{\sigma_x}, \frac{y_i}{\sigma_y}\right)$, where $\hat{\mathbf{x}}$ and $\hat{\mathbf{y}}$ denote the unit vectors along the $x$ and $y$ axes, respectively.

Next, we delve into this methodology in terms of two key questions:
\paragraph{Operational Principles of the Soft Floor Method}
As the number of optimization iterations increases, the value of $\gamma_{\mathcal{S}}$ rises, exacerbating the barrier between SLRs. 
Initially, when $\gamma_{\mathcal{S}} = 1$, as illustrated in \figRef{fig:soft_floor_example}, instances move easily between dies for global optimization. 
However, with $\gamma_{\mathcal{S}}$ increasing to 20, traversing between dies becomes more challenging and costly, directing the optimization towards refining local solutions.
Only instances located at the edges are considered to move between molds to obtain a more optimal solution.
By modulating the value of $\gamma_\mathcal{S}$, the algorithm strikes a balance between global and local searches, leading to better solutions in fewer iterations.
\paragraph{The Advancement of Our Method Over the Lifting Dimension Technique by \cite{analytical_place_3d_poisson}}
\begin{figure}[h]
	\centering
	\includegraphics[width=\columnwidth]{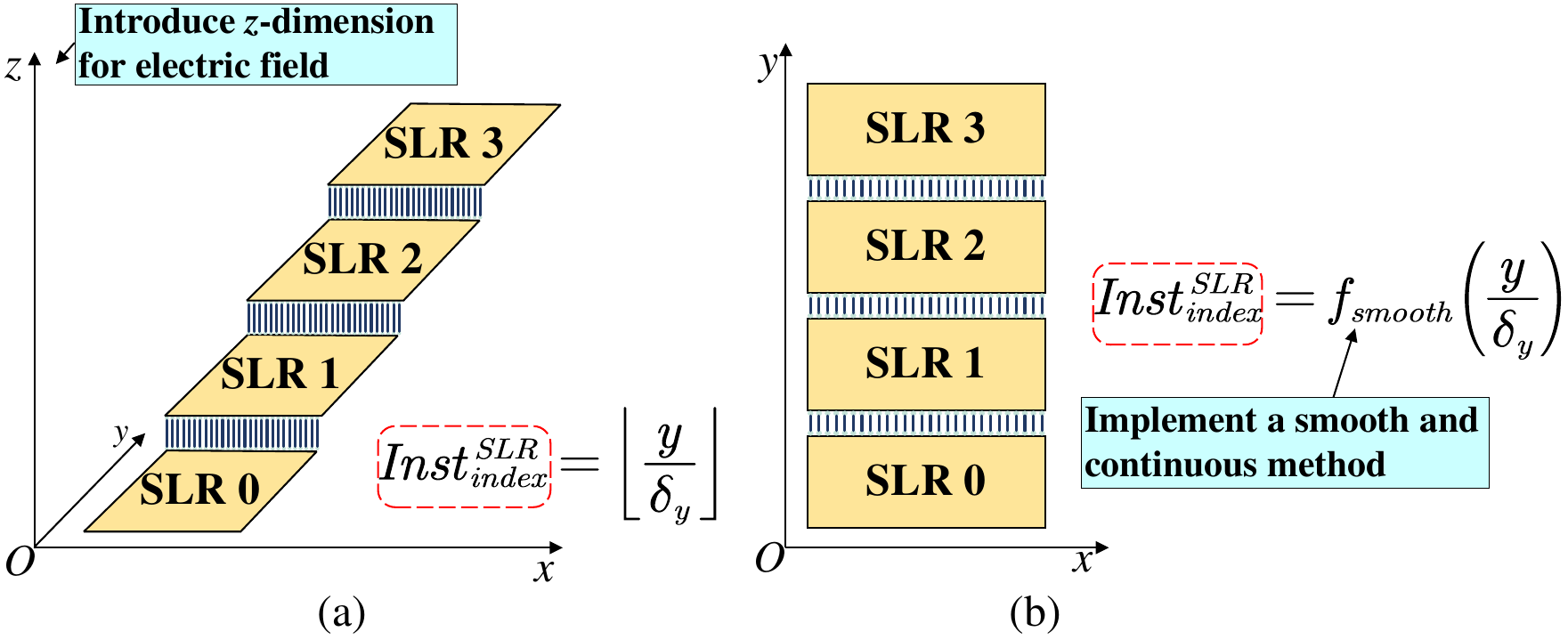}
	\caption{Comparison of the electric field modeling between the SOTA method~\cite{analytical_place_3d_poisson} and the proposed \texttt{LEAPS}.
		$Inst_{index}^{SLR}$ represents the SLR index of placeable instances.
		(a) The SOTA method utilizes the lifting dimension technique. 
		(b) The proposed \texttt{LEAPS} utilizes a smooth and continuous function $f_{smooth}$~(i.e., the soft floor method in~\secRef{sec:SoftFloorMethod}). 
	}
	\label{fig:Method_Comparison}
\end{figure}

While our soft floor method draws inspiration from the lifting dimension technique, it offers distinct improvements:
\begin{itemize}
	\item As shown in~\figRef{fig:Method_Comparison}\textcolor{blue}{(a)}, the lifting dimension technique in \cite{analytical_place_3d_poisson} specifically designed for $1\times4$ SLR topology and introduces an electric field dimension \( z \) with discrete SLR indexes, aiming to minimize wirelength in the \( z \)-direction using 3D Poisson equation and ADMM solver. However, this discrete approach leads to suboptimal results.
	\item In contrast, our soft floor method treats \( z \) as a continuous variable influenced by \( x \) and \( y \) coordinates, as depicted in~\figRef{fig:Method_Comparison}\textcolor{blue}{(b)}. 
	It can represent SLR indexes for any SLR topology.
	This continuous approach allows for smooth adjustments of instance coordinates, facilitating SLL minimization.
	Moreover, by employing the 2D Poisson equation, the placer simplifies computational demands and enhances design integration. 
\end{itemize}

In essence, the soft floor method provides a more adaptive and efficient approach to multi-die FPGA placement, promising optimal results. 
By transforming $\boldsymbol{z}$ into a continuous variable, it leverages gradient-based optimization, ensuring a differentiable wirelength model and improved placement results.

\subsection{Density Multiplier Updating for Multi-Die FPGA}
In FPGA placement, the density multiplier $\boldsymbol{\lambda}$ is pivotal for wirelength optimization, guiding the spreading rate of various resource types. 
While the method for updating $\boldsymbol{\lambda}$ has been extensively discussed in \cite{elfPlace_Meng}, our work introduces modifications tailored for multi-die FPGA, particularly considering SLL counts.

Our method initializes the density multiplier $\boldsymbol{\lambda}^{(0)}$ as follows:
\begin{equation}
	\boldsymbol{\lambda}^{(0)} = \eta \frac{\lVert \nabla \widetilde{W}_{\psi} \left(\boldsymbol{x}^{(0)},\boldsymbol{y}^{(0)}\right) \rVert_{1}}{\sum_{i \in V} q_i \lVert \boldsymbol{\xi}_{i}^{(0)} \rVert_{1}}\left(1, 1, \cdots, 1\right)^T.
	\label{eq:initial_lambda}
\end{equation}
The formula $\lVert \nabla \widetilde{W}_{\psi}(\cdot)\rVert_{1} = \lVert \nabla \widetilde{W}_{\mathcal{H}} (\cdot) + \psi \cdot
\nabla \widetilde{W}_{\mathcal{S}} (\cdot) \rVert_{1}$, distinct from \cite{elfPlace_Meng}, incorporates SLL counts into the initialization process. The initial placement location $(\boldsymbol{x}^{(0)}, \boldsymbol{y}^{(0)})$ and the initial electric field $\boldsymbol{\xi}_{i}^{(0)}$ of each instance are considered. The weight parameter $\eta$ and the L1 norm are calibrated to prioritize wirelength minimization in early iterations. We set $\eta$ to $10^{-4}$, applying uniform spreading weights across all resource types.

For the subsequent updating mechanism of $\boldsymbol{\lambda}$, we largely follow the subgradient update technique described in \cite{elfPlace_Meng}. 
This approach has been proven to enhance convergence efficiency and circuit quality.
Detailed technical aspects of this method are available in the cited work.

In conclusion, while our density multiplier updating mechanism builds upon the foundation set by \cite{elfPlace_Meng}, it introduces critical modifications to cater to the unique challenges posed by multi-die FPGA architecture, ensuring optimal placement results.

\subsection{Adaptive Wirelength-Weighting-Factor Adjusting}
\label{subsec.:Adaptive Wirelength-Weighting-Factor Adjusting}
To further improve the performance in solving the overall wirelength minimization problem, we also adaptively update the wirelength-weighting factor $\psi$ to balance the trade-off between HPWL minimization and SLL minimization. 
We apply an exponential moving average (EMA) and the Adam optimization algorithm, which has two main advantages: 1) improved convergence speed and 2) better trade-off between different objectives.

Defining the function $\mathcal{S}(x, y)$, which represents the number of SLLs. We first derive the growth of SLL counts, denoted by $\delta_{\mathcal{S}}^{(k+1)}$, in the $(k+1)$-th iteration. The calculation is performed as follows:
\begin{equation}
	\delta_{\mathcal{S}}^{(k+1)} = \mathcal{S}(\boldsymbol{x}^{(k+1)}, \boldsymbol{y}^{(k+1)}) - \mathcal{S}(\boldsymbol{x}^{(k)}, \boldsymbol{y}^{(k)}).
\end{equation}
This equation computes the change in the SLL counts from the $k$-th iteration to the $(k+1)$-th iteration, providing a quantitative measure of the SLL growth for the optimization process.

Next, we calculate the EMA of $\delta_{\mathcal{S}}^{k}$ using a weight parameter $\rho$, set to 0.9 for smooth convergence:
\begin{equation}
	E_{\mathcal{S}}^{(k+1)} = \rho \cdot \delta_{\mathcal{S}}^{(k+1)} + (1 - \rho) \cdot E_{\mathcal{S}}^{(k)}.
\end{equation}

We employ the Adam optimization algorithm to update $\psi$ based on the EMA value $E^{(k+1)}_{\mathcal{S}}$. 
The algorithm dynamically adjusts the learning rate using the first- and second-moment estimates of the gradient, with $E^{(k)}_{\mathcal{S}}$ serving as the gradient in this context. 
We compute the first-moment estimate $\psi_m$ and the second-moment estimate $\psi_v$:
\begin{equation}
	\psi_m = \beta_1 \cdot \psi_m + (1 - \beta_1) \cdot E_{\mathcal{S}}^{(k+1)},
	\label{eq:adam_psi_m}
\end{equation}

\begin{equation}
	\psi_v = \beta_2 \cdot \psi_v + (1 - \beta_2) \cdot (E_{\mathcal{S}}^{(k+1)})^2,
	\label{eq:adam_psi_v}
\end{equation}
Here, we set $\beta_1 = 0.9$ and $\beta_2 = 0.999$ as the exponential decay rates for the first- and second-moment estimates.
Then, we compute the bias-corrected first and second-moment estimates:

\begin{equation}
	\hat{\psi}_m = \frac{\psi_m}{1 - \beta_1},
	\label{eq:bias_corrected_psi_m}
\end{equation}

\begin{equation}
	\hat{\psi}_v = \frac{\psi_v}{1 - \beta_2},
	\label{eq:bias_corrected_psi_v}
\end{equation}

Lastly, we update $\psi^{(k)}$ with $\psi^{(k+1)}$ using the bias-corrected estimates:
\begin{equation}
	\psi^{(k+1)} = \psi^{(k)} + t_{\psi} \cdot \frac{\hat{\psi}_m}{\sqrt{\hat{\psi}_v} + \epsilon_{\psi}},
	\label{eq:update_psi}
\end{equation}
where $t_{\psi}$ is the step size, and $\epsilon_{\psi} = 10^{-8}$ is a small constant to prevent division by zero.

In essence, the use of EMA helps to reduce noise and maintain a good balance between HPWL and SLL counts. 
Not only that, the Adam optimization algorithm accelerates the convergence process and allows for an efficient trade-off between these objectives. 
This combined strategy features robustness to noise gradients and effective bias correction. 
It ensures an optimal balance between HPWL and SLL counts across the FPGA placement, improving overall performance.

\subsection{Improved Clock Network Planning Algorithm for Multi-Die FPGAs}
We introduce an advanced algorithm for clock network planning in multi-die FPGAs, aiming to satisfy clock constraints while effectively minimizing SLLs. 
This algorithm is structured into two key stages:
\begin{itemize}
	\item \textit{Instance-to-Clock-Region Mapping Generation with SLL Minimization}: This stage focuses on assigning instances to specific clock regions. 
	Our approach, inspired by the methods in \cite{direct_legalize}, extends beyond just adhering to clock routing constraints. 
	It integrates a novel optimization objective that concurrently addresses clock constraints and actively reduces SLL counts. See \secRef{subsubsec.: Instance-to-Clock-Region Mapping Generation with SLL Minimization}.
	\item \textit{The Advanced Clock Penalty}: In this stage, we incorporate a smooth, differentiable penalty function into the overall placement optimization. 
	This function is designed to subtly guide instances towards their specified clock regions, aligning with the clock network's layout and constraints. See \secRef{subsubsec.: The Advanced Clock Penalty}.
\end{itemize}
In our proposed algorithm, the synergy of these two stages results in a more balanced and efficient clock network planning for multi-die FPGAs.
It not only meets the critical clock constraints but also minimizes SLL counts, leading to an optimized placement and routing of the clock network.

\begin{table}[h]
	\centering
	\setlength{\tabcolsep}{6pt}
	\renewcommand{\arraystretch}{1.5}
	\caption{Symbols and Notions Used in Clock Network Planning.}
	\begin{tabular}[\columnwidth]{|c p{7.2cm}|}
		\hline
		$V$ 			& The set of instances. \\
		$S$ 			& The set of resource types. \\
		$V^{(s)}$ 	& The set of instances of resource type $s\in S$. \\
		$\mathcal{A}^{(s)}_{v}$ & The instance $v$'s demand for resource type $s \in S$. \\
		$\mathcal{R}$ 			& The set of clock regions. \\
		$C^{(s)}_r$ 			& The clock region $r$'s capacity for resource type $s \in S$.\\
		$D_{v,r}$ 				& The physical distance between instance $v$ and clock region $r$. \\
		$I_{v, r}$ 				& The increase in the number of SLLs if moving instance $v$ to
		clock region $r$. \\
		$\mathcal{E}$ 			& The set of clock nets. \\
		\hline
	\end{tabular}
	\label{cnp_symbols_and_notions}
\end{table}

\subsubsection{Instance-to-Clock-Region Mapping Generation with SLL Minimization}
\label{subsubsec.: Instance-to-Clock-Region Mapping Generation with SLL Minimization}
In this stage, our objective is to generate mappings from instances to clock regions. 
It requires satisfying clock constraints while minimizing the number of SLLs.
Initially, we introduce symbols and notions to clarify the problem, as shown in \tabRef{cnp_symbols_and_notions}.
Then, the instance-to-clock-region mapping process is formulated as a binary optimization problem, shown in Formulation (\ref{cnp_formulation}).
\begin{subequations}
	\begin{align}
		\underset{\boldsymbol{x}}{\text{minimize}} \quad 
		&\sum_{v \in V}\sum_{r \in \mathcal{R}} \left(D_{v,r} + \alpha I_{v, r}\right) \cdot \boldsymbol{x}_{v, r}, \label{eq:cnp_objective}\\
		\text{s.t.} \qquad 
		&\boldsymbol{x}_{v, r}\in\{0, 1\}, \forall v \in V, \forall r \in \mathcal{R}, \label{eq:cnp_x_status_constraint}\\
		&\sum_{r\in \mathcal{R}} \boldsymbol{x}_{v, r} = 1, \forall v \in V, \label{eq:cnp_x_sum_constraint}\\
		&\sum_{v \in \mathcal{V}} \mathcal{A}_{v}^{(s)} \cdot \boldsymbol{x}_{v, r} \leq C_{r}^{(s)}, \forall r \in \mathcal{R}, \forall s \in S, \label{eq:cnp_resource_constraint}\\
		&\text{Exist a legal clock routing w.r.t} \;\; \boldsymbol{x}. \label{eq:cnp_legal_clock_routing_constraint}
	\end{align}
	\label{cnp_formulation}
\end{subequations}

In the above formulation, the overall cost function (\eqRef{eq:cnp_objective}) is designed by summing 1) the physical distance $D_{v,r}$ between instances and clock regions and 2) the increase in SLL counts $I_{v, r}$. 
The two are weighed for importance by a factor $\alpha$.
The binary decision variable $\boldsymbol{x}_{v, r}$ (\eqRef{eq:cnp_x_status_constraint}) denotes the mapping status of instance $v$ to clock region $r$. 
The constraint (\eqRef{eq:cnp_x_sum_constraint}) specifies that each instance $v$ is mapped to exactly one clock region $r$.  
The upper limit of total resource demand per region is enforced by the constraint (\eqRef{eq:cnp_resource_constraint}), ensuring no clock region is overburdened.
Finally, the constraint (\eqRef{eq:cnp_legal_clock_routing_constraint}) ensures compliance with the legal clock routing and conforms to the constraints of the multi-die FPGA architecture. 
As such, the proposed formulation comprehensively explores the solution space, balancing the physical distance and the potential increase in SLL counts for clock network planning.

To solve this optimization problem~\eqRef{cnp_formulation}, we employ a \emph{branch-and-bound} based method proposed by~\cite{utplacef2x}, which advances the performance of clock-driven placement algorithms. 
This method utilizes a tree traversal-based heuristic to search for a huge solution space of possible variable assignments.

A critical distinction of our proposed method compared to previous works lies in its dual focus: not only does it minimize the total distance between instances and their designated clock regions, but it also crucially aims to reduce the overall increase in SLL counts. 
Addressing the SLL issue represents a notable advancement in this field. 
We introduce~\algRef{algorithm:compute_sll_improv} to address this challenge, detailing below its workflow for calculating the increase in SLL counts.
\paragraph{Streamlined Analysis of the Algorithm}
Initially, the increase in SLL counts $I_\mathcal{S}$ is set to zero, and the SLR index $cr_z$ for the clock region is computed, defining its location within the grid. 
Then, the algorithm checks each node in the candidate set. 
For each node, its SLR index is calculated and compared to the clock region’s index. 
If they match, the algorithm considers that there is no potential to increase the number of SLLs and proceeds to the next node. 
However, if the indexes differ, the algorithm further explores the node’s pins to identify eligible
nets to calculate the increase in the number of SLLs.
For each eligible net, the bounding box is computed, an updated bounding box is generated considering the instance-to-clock region assignment, and a partial increase in the number of SLLs is computed and added to $I_\mathcal{S}$. 
After evaluating all nodes, the algorithm concludes by returning the final total SLLs' increase $I_\mathcal{S}$. 
The algorithm is able to efficiently evaluate the potential increase in the number of SLLs while considering
the spatial relationships and interconnections of the nodes.

\begin{algorithm}[t]
	\caption{Calculation of the Increase in SLL Counts}
	\begin{algorithmic}[1]
		\Require{
			The set of candidate mapping nodes $N$, the target mapping clock region $cr$ with its central coordinates specified as $(cr_x, cr_y)$. The SLR's width $\delta_x$, the SLR's height $\delta_y$.
		}
		\Ensure{Total increase in SLL counts $I_\mathcal{S}$}
		\State $I_\mathcal{S} \gets 0$
		\State Compute $cr$'s SLR index $cr_z$ using \eqRef{eq:general_z_calculation}
		\ForAll{$n \in N$}
		\State Get node $n$'s coordinate $(n_x, n_y)$
		\State Get node $n$'s SLR index $n_z$ using \eqRef{eq:general_z_calculation}
		\If{$compare == 0$}
		\State \textbf{continue}
		\Else
		\State Get pins of node $n$ denoted as $P_n$
		\ForAll{$p_n$ in $P_n$}
		\State Get the net $e_p$ belonged to pin $p_n$
		\If{$e_p$ is not eligible}
		\State \textbf{continue}
		\EndIf
		\State Compute net $e_p$'s bounding box denoted as $B_e$
		\State Compute a partial increase in the number of SLLs $\Delta_\mathcal{S} \gets \lVert B_e - B_e^{'}\rVert_1$
		\State $I_\mathcal{S} \gets I_\mathcal{S} + \Delta_{\mathcal{S}}$
		\EndFor
		\EndIf
		\EndFor
		\State
		\Return {$I_\mathcal{S}$}
	\end{algorithmic}
	\label{algorithm:compute_sll_improv}
\end{algorithm}

\paragraph{Complexity Analysis of the Algorithm}
As for the algorithm’s complexity, the time complexity is mainly determined by the nested loops that iterates over the nodes and pins. 
Let $N$ represent the number of nodes, and $P$ denote the maximum number of pins per node. 
In the worst case, the algorithm needs to check all pins of all nodes, resulting in a time complexity of $O(NP)$. 
However, since only a small fraction of nets need to perform the calculation of the increased number of SLLs, the actual runtime is usually much less than the worst-case complexity. 
The space complexity is determined by the storage required for the data associated with nodes, pins, and various auxiliary data structures for intermediate computations. 
In general, the space complexity is proportional to the number of nodes, pins, and nets, making it $O(N + P + E)$, where $E$ represents the number of eligible nets.

\subsubsection{The Advanced Clock Penalty}
\label{subsubsec.: The Advanced Clock Penalty}
In the second stage, we implement an advanced clock penalty term to the placement objective for better adapting to the multi-die FPGA architecture while minimizing the number of SLLs and meeting clock constraints.

Unlike previous works~\cite{utplacef_2, clock-aware_ultrascale_fpga_placement}, which enforces a direct shift of instances to their clock regions. 
Instead, we adopt a novel gravitational attraction concept~\cite{clock_aware_place_Chen}, resembling a bowl-like pull, to guide instances toward their mapped clock regions.
The clock penalty function is expressed as: 
\begin{equation}
	\varGamma_i\left(\boldsymbol{x}_i, \boldsymbol{y}_i\right) = \varGamma_i\left(\boldsymbol{x}_i\right)^x+\varGamma_i\left(\boldsymbol{y}_i\right)^y.
\end{equation}

The penalty terms $\varGamma_i(\boldsymbol{x}_i)^x$ and $\varGamma_i(\boldsymbol{y}_i)^y$ correspond to the $x$ and $y$ directions, respectively.
Let $lo^x_i$, $hi^x_i$, $lo^y_i$, and $hi^y_i$ denote the left, right, bottom, and top boundary coordinates of the generated mapping result for instance $i$. 
We define $ \varGamma_i\left(\boldsymbol{x}_i\right)^x$ as,
\begin{equation}
	\varGamma_i\left(\boldsymbol{x}_i\right)^x = \left\{
	\begin{aligned}
		(\vec{x}_i - lo^x_i)^2,~& \vec{x}_i < lo^x_i, \\
		0, ~& lo^x_i \leq \vec{x}_i \leq hi^x_i, \\
		(\vec{x}_i - hi^x_i)^2, ~& hi^x_i < \vec{x}_i. \\
	\end{aligned}
	\right.
	\label{eq:clock_penalty}
\end{equation}
Here, $\varGamma(\boldsymbol{x}, \boldsymbol{y})$ denotes the sum of the clock penalty of all instances, i.e., $\varGamma(\boldsymbol{x}, \boldsymbol{y}) = \sum_{i \in \mathcal{V}}\varGamma_i\left(\boldsymbol{x}_i, \boldsymbol{y}_i\right)$.

The clock penalty multiplier $\eta$ is initially set to $0$. Upon resetting the clock penalty function $\varGamma(\cdot)$, $\eta$ is updated with the relative ratio between the gradient norms of the wirelength and the clock penalty to maintain the clock penalty function's stability.
\begin{equation}
	\eta = \frac{\iota \lVert\nabla \widetilde{W}_{\mathcal{\psi}}\rVert_2}{\lVert\nabla \varGamma \rVert_2 + \varepsilon}.
\end{equation}

As the placement optimization proceeds, the clock penalty multiplier $\eta$ is dynamically adjusted to balance the influence of wirelength and clock penalty terms in the objective function.
This adaptation ensures that the optimization algorithm maintains an appropriate focus on wirelength minimization
and compliance with clock region constraints.

After the instances are assigned to their respective clock regions, only about $1\%$ of instances are found outside their designated clock regions. 
Therefore, most of the instances do not incur any clock penalty.
Empirically, we set the parameters $\iota$ and $\varepsilon$ to $10^{-4}$ and $10^{-2}$, respectively.
This setting achieves an appropriate balance between the gradient norm ratio of wirelength and clock penalty terms.

As the optimization proceeds, instances that are still outside of their designated clock regions will be subjected to an increasing clock penalty. 
The size of the penalty will grow with the distance of the instance from its specified region, prompting the instance to move toward its specified clock region. 
This approach facilitates the smooth convergence of the optimization process while satisfying the clocking constraints imposed by the multi-die FPGA architecture.

In summary, this clock penalty method can dynamically adjust the clock penalty multiplier. 
This provides an efficient way to place instances in a multi-die FPGA architecture while minimizing the wirelength and satisfying clock region constraints. 
This approach enables improved placement quality and performance in comparison to existing methods, proving its applicability and effectiveness for modern FPGA designs.

\subsection{Clock- and SLL-aware Legalization \& Detailed Placement}
\label{subsec.:Clock- and SLL-aware Legalization & Detailed Placement}
Two critical constraints that need to be carefully considered during the legalization (LG) and detailed placement (DP) stages are clock feasibility and minimizing SLL counts. 
We delve into these constraints in the following discussion.

In the LG stage, we leverage the Direct Legalize (DL) algorithm~\cite{direct_legalize} to skillfully manage clock constraints. 
Due to the complexity of clock networks, modern FPGAs often introduce ``clock region constraints'' at this stage. 
To address this, we establish a legal clock-to-clock region assignment that specifies which cell can be positioned to which slice. 
Then, Our DL algorithm performs an additional check to discard cell-to-slice assignments that violate this assignment, thereby ensuring adherence to the clock region constraint.

In the DP stage, we leverage a clock-aware multi-stage ISM approach, drawing inspiration from the UTPlaceF series~\cite{utplacef, utplacef_2}. 
The approach utilizes an iterative minimum-cost-flow-based cell assignment technique to optimize wirelength and routability while adhering to complex clock constraints, resulting in clock-legal and high-quality placement solutions.

Addressing the SLL minimization challenge involves estimating the potential increase in SLL counts due to instance relocations during legalization and detailed placement.
This estimation is seamlessly incorporated into the optimization objectives, mirroring the~\algRef{algorithm:compute_sll_improv} applied in clock network planning. 
The goal is to ensure that the overall optimization objective is minimized.

To concretely demonstrate our methodology, we consider the SLL optimization in the LG to illustrate the practical details.
Given the DL algorithm concurrently explores the solution spaces of placement and packing. 
This requires a scoring function that encapsulates both placement- and packing-related metrics. 
Given a slice $s$ and a cluster $c$, the score of $c$ in $s$ is defined as follows:

\begin{equation}
	\begin{aligned}
		\text{SCORE}(c,s) &= \sum_{e\in \mathcal{E}(c)}\frac{\text{InternalPins}(e, c) - 1}{\text{TotalPins}(e) - 1} \\
		&- \varphi (\Delta HPWL(c, s) + \alpha_\text{LG}\Delta SLL(c, s))
	\end{aligned}
\end{equation}
Here, $\mathcal{E}(c)$ denotes the set of nets with at least one cell in $c$, $\text{TotalPins}(e)$ represents the total pin count of net $e$, $\text{InternalPins}(e, c)$ indicates the number of pins of net $e$ in $c$, and $\Delta HPWL(c, s)$ and $\Delta SLL(c, s)$ denote the increase in HPWL and the number of SLLs when moving cells in $c$ from their flat initial placement (FIP) locations to $s$. 
The positive weighting parameters $\varphi$ and $\alpha_{\text{LG}}$ are empirically set to $0.02$ and $4.0$, respectively. 
The first term defines the clustering score, granting higher scores to clusters that convert more external nets into internal ones, effectively reducing routing demands and enhancing routability. 
The second term favors candidates that significantly reduce the wirelength and the number of SLLs.

The SLL optimization in the DP is consistent with the approach in the LG above, while also taking clock constraints into account. 
This ensures the overall optimization goal, including SLL minimization and clock routing constraints, guarantees high-quality placement results. 
This process is not further elaborated here and can be referred to in the description in the LG.

\renewcommand{\arraystretch}{1.1}
\begin{table*}[t]
	\centering
	\caption{Comparison of Super Long Lines (×$10^0$), Half-Perimeter Wirelength (×$10^3$), and Runtime (Seconds) for Multi-Die FPGA with 1$\times$4 SLR Topology on ISPD 2017 Benchmarks.}
	\label{tab:sota_comparison_ispd2017}
	\resizebox{\textwidth}{!} {
		\begin{tabular}{|c|cc|ccc|ccc|ccc|cccc|}
			\hline
			\multirow{2}{*}{Design} & \multirow{2}{*}{\#LUT/\#FF/\#BRAM/\#DSP} & \multirow{2}{*}{\#Clock} & \multicolumn{3}{c|}{\texttt{ICCAD'17}~\cite{clock_aware_place_Kuo}} & \multicolumn{3}{c|}{\texttt{Min-cut + ICCAD'17}~\cite{clock_aware_place_Kuo}} & \multicolumn{3}{c|}{\texttt{ICCAD'19}~\cite{analytical_place_3d_poisson}} & \multicolumn{4}{c|}{The Proposed \texttt{LEAPS}} \\
			& & & SLL		& HPWL		& CRT (s)		& SLL		& HPWL		& CRT (s)	& SLL		& HPWL		& CRT (s)		& SLL		& HPWL	& CRT(s) & GRT(s)
			\\ 
			\hline \hline
			
			\texttt{CLK-FGPA01}              
			& 211K/324K/164/75 & 32                  & 19707           & 1933691 & 2939  & 15039 & 2126497 & 7963  & 14817 & 1916227 & 3227  & \textbf{4873}  & \textbf{1658361} & 1697 & \textbf{123}  \\
			
			\texttt{CLK-FGPA02}              
			& 230K/280K/236/112                      & 35                       & 19245            & 1949266             & 3356			& 14937            & 2138430		& 7772  & 14470 & 1927038 & 3225  & \textbf{7192}  & \textbf{1777341} & 1552 & \textbf{121} \\
			
			\texttt{CLK-FGPA03}              
			& 410K/481K/850/395                     & 57                       
			& 33915            & 4760837            & 7410          
			& 24310            & 5702452  			& 17545 & 22500 & 4688170 & 7251  & \textbf{14285} & \textbf{4487928} & 2377 & \textbf{202}\\
			
			\texttt{CLK-FGPA04}              
			& 309K/372K/467/224                     & 44                       
			& 22774            & 3388240            & 6015           
			& 17317            & 4163495  			& 13060 & 17123 & 3389653 & 5419  & \textbf{10852} & \textbf{3094173} & 2148 & \textbf{148} \\
			
			\texttt{CLK-FGPA05}              
			& 393K/469K/798/150                     & 56                       
			& 28246            & 4147683            & 7460           
			& 21745            & 5112935  			& 17533 & 21238 & 4066860 & 7275  & \textbf{11777} & \textbf{3821386} & 2269 & \textbf{189}    \\
			
			\texttt{CLK-FGPA06}              
			& 425K/511K/872/420                     & 58          
			& 30526            & 5007798            & 8261           
			& 21260            & 6128113  			& 12708 & 20988 & 5152846 & 5686  & \textbf{15959} & \textbf{4625645} & 2355 & \textbf{214}    \\
			
			\texttt{CLK-FGPA07}              
			& 254K/309K/313/149                      & 38                       
			& 14916            & 2096178             & 3747           
			& 11079            & 2271849  			 & 8582  & 11215 & 2047259 & 3561  & \textbf{6813}  & \textbf{1905011}  & 1688 & \textbf{127}    \\
			
			\texttt{CLK-FGPA08}              
			& 212K/257K/161/75                       & 32                       
			& 16711            & 1673570             & 2812           
			& 13457            & 2143600  			 & 8401  & 12565 & 1661350 & 3509  & \textbf{4849}  & \textbf{1545018}  & 1554 & \textbf{109}    \\
			
			\texttt{CLK-FGPA09}              
			& 231K/358K/236/112                      & 35                       
			& 16275            & 2162916             & 3994           
			& 10282            & 2836349  		     & 10879 & 10485 & 2177478 & 4512  & \textbf{6508}  & \textbf{1891086}  & 1816 & \textbf{131}    \\
			
			\texttt{CLK-FGPA10}              
			& 327K/506K/542/255                      & 47                       
			& 22584            & 3886385             & 6396           
			& 17793            & 4716132  		     & 18464 & 17233 & 3970566 & 7675  & \textbf{13816} & \textbf{3301351}  & 2273 & \textbf{180}    \\
			
			\texttt{CLK-FGPA11}              
			& 300K/468K/454/224                     & 44                       
			& 26024            & 3676642            & 6339           
			& 19356            & 4573412  			& 15325 & 19567 & 3697769 & 6359  
			& \textbf{11052} & \textbf{3138093}  & 2731 & \textbf{165}    \\
			
			\texttt{CLK-FGPA12}              
			& 277K/430K/389/187                      & 41                       
			& 25683            & 2814733             & 4703           
			& 19275            & 3109834  			 & 12768 & 18559 & 2811424 & 5702  & \textbf{10533} & \textbf{2453370} & 1916 & \textbf{151}    \\
			
			\texttt{CLK-FGPA13} 
			& 339K/405K/570/262 					 & 47 
			& 32248 		   & 3464495 			 & 4750 
			& 24774 		   & 4297976 		     & 14354 & 24999 & 3422521 			& 5956  & \textbf{10003} & \textbf{3172093} & 2057 & \textbf{160}    \\
			\hline \hline
			\multicolumn{3}{|c|}{Norm.\footnotemark[1]} & 2.403 & 1.111 & 33.753 & 1.795 & 1.338 & 81.858 & 1.757 & 1.110 & 34.335 & \textbf{1.000} & \textbf{1.000} & 13.084 & \textbf{1.000}
			\\ 
			\hline
		\end{tabular}
	}
\end{table*}

\begin{table*}[t]
	\centering
	\caption{HPWL and SLL Evaluations With Different Stages Optimizations on ISPD 2017 Benchmarks.}
	\resizebox{\textwidth}{!} {
		\begin{tabular}{|c|cc|cc|cc|cc|cc|cc|}
			\hline
			\multirow{3}{*}{Design} & \multicolumn{6}{c|}{$1\times4$ SLR Topology} & \multicolumn{6}{c|}{$2\times2$ SLR Topology} \\
			\cline{2-13}
			& \multicolumn{2}{c|}{\texttt{LEAPS(GP)}} & \multicolumn{2}{c|}{\texttt{LEAPS(GP+LG+DP)}} & \multicolumn{2}{c|}{\texttt{LEAPS(GP+LG+DP+CNP)}} & \multicolumn{2}{c|}{\texttt{LEAPS(GP)}} & \multicolumn{2}{c|}{\texttt{LEAPS(GP+LG+DP)}} & \multicolumn{2}{c|}{\texttt{LEAPS(GP+LG+DP+CNP)}} \\
			& \multicolumn{1}{c}{SLL} & HPWL  & \multicolumn{1}{c}{SLL} & HPWL  & \multicolumn{1}{c}{SLL} & HPWL  & \multicolumn{1}{c}{SLL} & HPWL  & \multicolumn{1}{c}{SLL} & HPWL  & \multicolumn{1}{c}{SLL} & \multicolumn{1}{c|}{HPWL} \\
			\hline \hline			
			\texttt{CLK-FPGA01} & 5180  & 1658842 & 4916  & 1659305 & \textbf{4873}  & \textbf{1658361} & 10598 & \textbf{1629348} & 10173 & 1631056 & \textbf{10026} & 1631274 \\
			\texttt{CLK-FPGA02} & 7523  & 1800437 & 7278  & 1799139 & \textbf{7192}  & \textbf{1777341} & 14480 & 1788592 & \textbf{13928} & \textbf{1773191} & 13984 & 1791406 \\
			\texttt{CLK-FPGA03} & 14833 & 4495993 & 14430 & 4494080 & \textbf{14285} & \textbf{4487928} & 24088 & \textbf{4471987} & 23626 & 4475201 & \textbf{23371} & 4474117 \\
			\texttt{CLK-FPGA04} & 11034 & \textbf{3090311} & \textbf{10845} & 3093255 & 10852 & 3094173 & 19781 & \textbf{3068101} & \textbf{19014} & 3070280 & 19131 & 3068247 \\
			\texttt{CLK-FPGA05} & 12079 & 3832208 & \textbf{11773} & 3832323 & 11777 & \textbf{3821386} & 21617 & \textbf{3843644} & 20799 & 3843667 & \textbf{20735} & 3844829 \\
			\texttt{CLK-FPGA06} & 16351 & \textbf{4622926} & 15983 & 4624631 & \textbf{15959} & 4625645 & 26867 & \textbf{4642177} & 26124 & 4648913 & \textbf{25970} & 4642732 \\
			\texttt{CLK-FPGA07} & 7234  & 1912074 & 6936  & 1912016 & \textbf{6813}  & \textbf{1905011} & 12150 & \textbf{1912148} & \textbf{11494} & 1915191 & 11584 & 1913722 \\
			\texttt{CLK-FPGA08} & 4915  & 1545356 & \textbf{4588}  & \textbf{1544583} & 4849  & 1545018 & 10182 & 1530010 & \textbf{9595}  & 1531410 & 9706  & \textbf{1524861} \\
			\texttt{CLK-FPGA09} & 6864  & \textbf{1889271} & 6526  & 1890708 & \textbf{6508}  & 1891086 & 10922 & 1904842 & 10388 & 1904557 & \textbf{10339} & \textbf{1904389} \\
			\texttt{CLK-FPGA10} & 14285 & \textbf{3299965} & 13941 & 3304380 & \textbf{13816} & 3301351 & 21470 & \textbf{3304337} & 20890 & 3305586 & \textbf{20717} & 3306488 \\
			\texttt{CLK-FPGA11} & 11042 & \textbf{3136026} & \textbf{10690} & 3136950 & 11052 & 3138093 & 16921 & \textbf{3130795} & 16227 & 3134114 & \textbf{16087} & 3131357 \\
			\texttt{CLK-FPGA12} & 10985 & \textbf{2451674} & 10825 & 2452170 & \textbf{10533} & 2453370 & 15244 & 2460003 & 14723 & 2461136 & \textbf{14593} & \textbf{2460000} \\
			\texttt{CLK-FPGA13} & 10352 & 3177396 & 10172 & 3181802 & \textbf{10003} & \textbf{3172093} & 18742 & \textbf{3168715} & 18196 & 3170607 & \textbf{18179} & 3172810 \\
			\hline\hline
		    Norm.  & 1.0324 & 1.0011 & 1.0030 & 1.0015 & \textbf{1.0000} & \textbf{1.0000} & 1.0403 & \textbf{0.9997} & 1.0035 & 1.0000 & \textbf{1.0000} & 1.0000 \\
			\hline
		\end{tabular}%
		\label{tab:sll&hpwl_comparsion_with_different_stages}%
	}
\end{table*}

\section{Experimental Results}
\label{sec:experimental_results}
\subsection{Comparison with the SOTA methods}
We implemented our GPU-accelerated placer in C++ and Python along with the open-source machine learning framework PyTorch for fast gradient back-propagation.
We conduct experiments on a Ubuntu 22.04 LTS platform that consists of an Intel(R) Xeon(R) Gold 6248 CPU @ 3.00GHz (24 cores), an NVIDIA RTX3090 GPU, and 128GB memory. 

To comprehensively compare our \texttt{LEAPS} with other SOTA placers, we evaluated its performance using the \textit{ISPD 2017 benchmarks}, specifically targeting multi-die FPGA with a 1×4 SLR topology. 
These evaluations focus on three key metrics: minimization of super long lines (SLL), optimization of half-perimeter wirelength (HPWL), and overall runtime efficiency. 
Notably, we further dissect the runtime into CPU runtime (CRT) and GPU runtime (GRT) to highlight the GPU acceleration capabilities of our placement method.

The characteristics and comparative analysis of various FPGA placement algorithms, including \texttt{ICCAD'17}~\cite{clock_aware_place_Kuo}, \texttt{Min-cut + ICCAD'17}~\cite{clock_aware_place_Kuo}, \texttt{ICCAD'19}~\cite{analytical_place_3d_poisson}, and our proposed \texttt{LEAPS}, are detailed in Table~\ref{tab:sota_comparison_ispd2017}, focusing on the \emph{ISPD 2017 contest benchmark}. The rationale behind selecting these specific algorithms for comparison is as follows:
\begin{itemize}
	\item \texttt{ICCAD'17}~\cite{clock_aware_place_Kuo} and \texttt{Min-cut + ICCAD'17}~\cite{clock_aware_place_Kuo} are included despite \texttt{ICCAD'17}~\cite{clock_aware_place_Kuo}  not being a multi-die FPGA placer. 
	It represents a significant clock-aware placement algorithm. 
	The \texttt{Min-cut + ICCAD'17} setup, which combines the Min-cut method with \texttt{ICCAD'17}~\cite{clock_aware_place_Kuo}, not only provides a balanced comparison but is also pivotal in the \texttt{ICCAD'19}~\cite{analytical_place_3d_poisson} analysis, serving as a benchmark method.
	This method divides blocks into four subsets for placement within each die, providing a unique approach to FPGA placement.
	\item \texttt{ICCAD'19}~\cite{analytical_place_3d_poisson} is considered for its recent advancements as a state-of-the-art (SOTA) method, particularly addressing SLL challenges with clock- and SLL-aware techniques.
\end{itemize}
Notably, recent heterogeneous FPGA placement algorithms such as \texttt{elfPlace}~\cite{elfPlace_Meng}, \texttt{DREAMPlaceFPGA}~\cite{dreamplacefpga}, \texttt{AMF-Placer}~\cite{amfplacer} are excluded from this comparison. 
The key reason for their exclusion is the omission of clock constraints in these algorithms, a detail underscored in~\secRef{sec:Comparative Features of LEAPS and Other Placers}.
The assessment in \tabRef{tab:FPGAPlacers} is based on three key features: handling clock constraints, supporting multi-die architecture, and leveraging GPU acceleration. 
In these aspects, \texttt{LEAPS} demonstrates its powerful capabilities in meeting the demands of modern FPGA design, distinguishing itself in the field of FPGA placement.
This distinction is crucial, as overlooking clock constraints can significantly affect wirelength metrics post-placement, leading to an inaccurate comparison of performance metrics.
Additionally, \texttt{OpenPARF}\cite{openparf}, representing our preliminary work, is not compared directly. 
However, the superiority of the \texttt{LEAPS} framework is evident from the results presented in Tables~\ref{tab:sota_comparison_ispd2017},~\ref{tab:sll&hpwl_comparsion_with_different_stages}, and~\ref{tab:wlw_experiment}.
For enhanced clarity and emphasis, the most superior results in these tables are highlighted in bold.
	
\footnotetext[1]{
	The Norm. in this table are calculated using the relative improvement method. 
	This differs from the relative reduction percentage used in the main text, leading to variations in the reported values.
}

The analysis presented in Table~\ref{tab:sota_comparison_ispd2017} clearly indicates that our \texttt{LEAPS} method surpasses other algorithms across all evaluated metrics, achieving notably lower counts of super long lines (SLL) and improved half-perimeter wirelength (HPWL) for all benchmark designs. Additionally, \texttt{LEAPS} demonstrates a substantial advantage in runtime, consistently completing placements more rapidly than its counterparts. This remarkable enhancement in performance is largely due to the method's efficient optimization techniques and the integration of GPU acceleration.
It's noteworthy that even when the GPU acceleration factor is set aside, the CPU-based implementation of \texttt{LEAPS} still significantly outpaces the current state-of-the-art, \texttt{ICCAD'19}~\cite{analytical_place_3d_poisson}, with an approximate 2.62$\times$ speedup in runtime. 
In comparison to the latest SOTA method \texttt{ICCAD’19}~\cite{analytical_place_3d_poisson}, \texttt{LEAPS} with GPU acceleration demonstrates a substantial reduction in SLL by 43.08\% and in HPWL by 9.99\%, along with a significant 34.335$\times$ speedup in runtime.
These results underscore \texttt{LEAPS}'s ability to achieve more optimal placements with lower computational demands.

In conclusion, \texttt{LEAPS} demonstrates clear superiority over other algorithms in SLL, HPWL, and runtime metrics for the ISPD 2017 benchmarks. 
Its combination of efficient optimization and GPU acceleration not only minimizes HPWL and SLL counts but also reduces computational overhead, making it an effective solution for multi-die FPGA placement.

\subsection{Effectiveness Validation of Optimization Techniques}
In this section, we conduct a thorough validation of the techniques presented in the \texttt{LEAPS} framework, specifically tailored for multi-die FPGA placement.
To achieve this, we design two sets of experiments: the first evaluates the impact of optimizing SLL at different stages of the placement process, while the second assesses the adaptive wirelength-weighting-factor adjusting method (hereafter referred to as the WLW method) in the GP, which enables trade-offs between HPWL and SLL counts.
These experiments aim to provide a comprehensive understanding of how each technique within \texttt{LEAPS} contributes to the overall placement efficacy.

\subsubsection{Necessity of Full-flow Optimization in LEAPS}
Our primary focus is on the full-flow optimization of the number of SLLs, driven by the premise that SLL minimization should be a continuous effort throughout the entire placement process, not limited to the GP stage alone.
We conducted comparative experiments, as detailed in \tabRef{tab:sll&hpwl_comparsion_with_different_stages}, evaluating HPWL and SLL across three scenarios: 1) optimization solely during the GP stage (abbreviated as \texttt{LEAPS(GP)}), 2) optimization across the GP, LG, and DP stages (abbreviated as \texttt{LEAPS(GP+LG+DP)}), and 3) optimization extending into the clock network planning (CNP) stage (abbreviated as \texttt{LEAPS(GP+LG+DP+CNP)}).

Results from experiments using both $1\times4$ and $2\times2$ SLR topologies on the ISPD 2017 benchmark distinctly demonstrate significant reductions in SLL counts and improvements in wirelength optimization.
Importantly, the application of optimization at the LG, DP, and CNP stages leads to a progressive decrease in SLL counts, confirming their effectiveness in refining SLL optimization in multi-die FPGA designs.
Although optimization in the LG, DP, and CNP typically results in a minor increase in HPWL within acceptable limits, they occasionally produce a decrease in HPWL. 
This enhancement may be attributed to more refined clock network planning and wirelength objective, and is also likely influenced by inherent coupling mechanisms within these topologies.
However, it is acknowledged that our current understanding of these blind spots is incomplete, prompting further investigation. This area forms the nucleus of our ongoing research endeavors.
Moreover, our analysis suggests that compared to the $2\times2$ SLR topology, the $1\times4$ configuration results in fewer SLLs while maintaining comparable HPWL.
This suggests a potential preference for the $1\times4$ topology in multi-die FPGAs with four SLRs. 
However, a deeper study into other design performance aspects, like timing and routed wirelength, is essential for a definitive finding.

Upon the normalized data, it becomes evident that the comprehensive \texttt{LEAPS} framework (encompassing GP, LG, DP, and CNP) is the most effective, significantly enhancing SLL and HPWL performance in the evaluated designs.
These results not only demonstrate the efficacy of the \texttt{LEAPS} framework but also validate our strategic approach towards optimizing the entire workflow in multi-die FPGA design.

\subsubsection{Effectiveness of Adaptive Wirelength-weighting-factor Adjusting Method}
\begin{table}[htbp]
	
	\centering
	\caption{Comparative Performance Analysis of the \texttt{LEAPS} Framework Utilizing Versus Omitting the WLW Method in a 1×4 SLR Topology.}
	\resizebox{\columnwidth}{!} {
	\begin{tabular}{|c|cccc|}
			\hline
			\multirow{2}{*}{Design} & \multicolumn{2}{c|}{\texttt{LEAPS(without WLW)}} & \multicolumn{2}{c|}{\texttt{LEAPS(with WLW)}} \\
			\multicolumn{1}{|c|}{} & \multicolumn{1}{c}{SLL} & \multicolumn{1}{c|}{HPWL} & \multicolumn{1}{c}{SLL} & \multicolumn{1}{c|}{HPWL} \\
			\hline \hline
			\texttt{CLK-FPGA01} & 4992  & \multicolumn{1}{c|}{\textbf{1654111}} & \textbf{4873}  & 1658361 \\
			\texttt{CLK-FPGA02} & 7425  & \multicolumn{1}{c|}{1777650} & \textbf{7192}  & \textbf{1777341} \\
			\texttt{CLK-FPGA03} & 14971 & \multicolumn{1}{c|}{\textbf{4485711}} & \textbf{14285} & 4487928 \\
			\texttt{CLK-FPGA04} & 11989 & \multicolumn{1}{c|}{\textbf{3089971}} & \textbf{10852} & 3094173 \\
			\texttt{CLK-FPGA05} & 12942 & \multicolumn{1}{c|}{3826220} & \textbf{11777} & \textbf{3821386} \\
			\texttt{CLK-FPGA06} & 16465 & \multicolumn{1}{c|}{4627091} & \textbf{15959} & \textbf{4625645} \\
			\texttt{CLK-FPGA07} & 6942  & \multicolumn{1}{c|}{\textbf{1902767}} & \textbf{6813}  & 1905011 \\
			\texttt{CLK-FPGA08} & 5532  & \multicolumn{1}{c|}{1546489} & \textbf{4849}  & \textbf{1545018} \\
			\texttt{CLK-FPGA09} & 6679  & \multicolumn{1}{c|}{\textbf{1889111}} & \textbf{6508}  & 1891086 \\
			\texttt{CLK-FPGA10} & 14017 & \multicolumn{1}{c|}{3301502} & \textbf{13816} & \textbf{3301351} \\
			\texttt{CLK-FPGA11} & 11711 & \multicolumn{1}{c|}{\textbf{3126103}} & \textbf{11052} & 3138093 \\
			\texttt{CLK-FPGA12} & 10846 & \multicolumn{1}{c|}{\textbf{2447167}} & \textbf{10533} & 2453370 \\
			\texttt{CLK-FPGA13} & 10215 & \multicolumn{1}{c|}{3173383} & \textbf{10003} & \textbf{3172093} \\
			\hline \hline
			Norm.  & 1.048 & \multicolumn{1}{c|}{0.999} & \textbf{1.000} & \textbf{1.000} \\
			\hline
		\end{tabular}%
	}
	\label{tab:wlw_experiment}%
\end{table}

By integrating SLL counts into the conventional wirelength objective, the \texttt{LEAPS} framework innovates with the WLW method.
This method aims to strike a balance between HPWL and SLL counts during the GP, addressing one of the key challenges in \texttt{LEAPS}. 
\tabRef{tab:wlw_experiment} presents a comparative analysis of \texttt{LEAPS} with the WLW method (\texttt{LEAPS(with WLW)}) and without the WLW method (\texttt{LEAPS(without WLW)}), illustrating the method's effectiveness in reducing SLL counts with a minimal impact on HPWL. Specifically, in the $1\times4$ SLR topology, the WLW method achieved a notable 4.58\% reduction in SLLs with only a marginal 0.1\% increase in HPWL.
These results validate the WLW method's efficacy in achieving a delicate balance between minimizing SLL counts and maintaining HPWL, underscoring its importance in the \texttt{LEAPS} framework for multi-die FPGA placement.

\section{Conclusion}
\label{sec:conclusion}
In this paper, we have introduced \texttt{LEAPS}, a comprehensive and adaptable multi-die FPGA placement algorithm that addresses the challenges of minimizing SLL counts while optimizing essential design constraints, such as wirelength, routability, and clock routing. 
Our key contributions include a high-performance nested optimization algorithm with adaptive wirelength-weighting-factor adjusting, a soft floor method for handling
any multi-die FPGA SLR topology, and the continuous optimization of SLLs throughout the entire placement process,
including LG and DP stages.
Experimental results demonstrate that our method significantly outperforms the SOTA algorithm, achieving an average reduction of 43.08\% and 9.99\% in SLL counts and HPWL, respectively, and a 34.34$\times$ speedup in execution efficiency.

Future research may involve refining the \texttt{LEAPS} framework by developing advanced optimization techniques or employing machine learning to learn from previous placement experiences.
Additionally, integrating our algorithm with other placement and routing tools could enhance seamless interoperability and collaboration between different stages of the FPGA design flow. 
In conclusion, \texttt{LEAPS} offers a promising foundation for addressing challenges in multi-die FPGA placement, setting the stage for future advancements in this field and contributing to the ongoing development of high-performance computing systems.

\end{document}